\documentclass[english,citesort,aps,showpacs,showkeys,preprintnumbers,prd,floatfix,nofootinbib,superscriptaddress,notitlepage,twocolumn]{revtex4-1}

\usepackage{graphicx}
\usepackage{nicefrac}
\usepackage{lineno}
\usepackage{amsmath}
\usepackage{amsfonts}
\usepackage{hyperref}
\usepackage{subfig}
\usepackage{color}
\usepackage{enumitem}
\usepackage{multirow}
\usepackage{array}
\usepackage{lipsum}
\usepackage{ulem}
\usepackage[labelfont=bf]{caption}
\def\note#1{\vspace{6pt}\noindent{\bf #1:}}

\def\ncteqpp{{\tt \textbf{nCTEQ++}}}
\def\ncteqwz{{\tt \textbf{nCTEQ15WZ}}}
\def\wz{$W^\pm\!/Z$}
\def\chidof{\chi^2/N_{dof}}

\definecolor{purple}{RGB}{128,0,128}

\begin{document}
\modulolinenumbers[5]
\setlength\linenumbersep{2pt}

\preprint{
  IFJPAN-IV-2020-4, \ 
    KA-TP-07-2020, \ 
    MS-TP-20-27, 
}
\preprint{
   P3H-20-035, \ 
    SMU-HEP-20-04
}
\null \vspace{0.2cm}

\title{Impact of LHC vector boson production in heavy ion collisions on strange PDFs}

\author{A. Kusina}
\email{Aleksander.Kusina@ifj.edu.pl}
\affiliation{Institute of Nuclear Physics Polish Academy of Sciences,\\ PL-31342 Krakow, Poland}

\author{T.~Je\v{z}o}
\email{tomas.jezo@kit.edu}
\affiliation{Institute for Theoretical Physics, KIT, Karlsruhe, Germany}

\author{D.~B.~Clark}
\affiliation{Southern Methodist University, Dallas, TX 75275, USA }

\author{P.~Duwent\"aster}
\affiliation{Institut f{\"u}r Theoretische Physik, Westf{\"a}lische Wilhelms-Universit{\"a}t
M{\"u}nster, Wilhelm-Klemm-Stra{\ss}e 9, D-48149 M{\"u}nster, Germany }

\author{E.~Godat}
\affiliation{Southern Methodist University, Dallas, TX 75275, USA }

\author{T.~J.~Hobbs}
\affiliation{Southern Methodist University, Dallas, TX 75275, USA }
\affiliation{Jefferson Lab, EIC Center, Newport News, VA 23606, USA}

\author{J.~Kent}
\affiliation{Southern Methodist University, Dallas, TX 75275, USA }

\author{M.~Klasen}
\affiliation{Institut f{\"u}r Theoretische Physik, Westf{\"a}lische Wilhelms-Universit{\"a}t
M{\"u}nster, Wilhelm-Klemm-Stra{\ss}e 9, D-48149 M{\"u}nster, Germany }

\author{K.~Kova\v{r}\'{i}k}
\email{karol.kovarik@uni-muenster.de}
\affiliation{Institut f{\"u}r Theoretische Physik, Westf{\"a}lische Wilhelms-Universit{\"a}t
M{\"u}nster, Wilhelm-Klemm-Stra{\ss}e 9, D-48149 M{\"u}nster, Germany }

\author{F.~Lyonnet}
\affiliation{Southern Methodist University, Dallas, TX 75275, USA }

\author{K.~F.~Muzakka}
\affiliation{Institut f{\"u}r Theoretische Physik, Westf{\"a}lische Wilhelms-Universit{\"a}t
M{\"u}nster, Wilhelm-Klemm-Stra{\ss}e 9, D-48149 M{\"u}nster, Germany }

\author{F.~I.~Olness}
\email{olness@smu.edu}
\affiliation{Southern Methodist University, Dallas, TX 75275, USA }

\author{I.~Schienbein}
\email{ingo.schienbein@lpsc.in2p3.fr}
\affiliation{Laboratoire de Physique Subatomique et de Cosmologie, Université
Grenoble-Alpes, CNRS/IN2P3, 53 avenue des Martyrs, 38026 Grenoble,
France }

\author{J.~Y.~Yu}
\affiliation{Southern Methodist University, Dallas, TX 75275, USA }

\begin{abstract}
The extraction of the strange quark parton distribution function (PDF)
poses a long-standing puzzle. 
Measurements from neutrino-nucleus  deep inelastic scattering (DIS)
experiments suggest the strange quark
is suppressed compared to the light sea quarks, while recent studies
of \wz\ boson production at the LHC imply a larger strange component
at small~$x$ values. 
As the parton flavor determination in the proton depends on nuclear
corrections, e.g. from heavy-target DIS, LHC heavy ion measurements can
provide a distinct perspective to help clarify this situation. 
In this investigation we  extend the nCTEQ15 nPDFs to study the impact of the
LHC proton-lead \wz\ production data on both the flavor differentiation
and nuclear corrections.
This complementary data set provides new insights on both the LHC
\wz\  proton analyses and the neutrino-nucleus DIS data.
We identify these new nPDFs as \ncteqwz.
Our calculations are performed using a new implementation of the nCTEQ
code (\ncteqpp) based on C++ which enables us to easily interface to
external programs such as HOPPET, APPLgrid and MCFM.
Our results indicate that, as suggested by the proton data, the
small~$x$ nuclear strange sea appears larger than previously expected,
even when the normalization of the $W^{\pm}/Z$ data is accommodated in
the fit.
Extending the nCTEQ15 analysis to include  LHC \wz\ data represents
an important step as we advance toward the next generation of nPDFs. 
\end{abstract}

\date{\today}

\maketitle
\null \vspace{-1.2cm}
\tableofcontents{}
\def\tabData{
\begin{table}[t]
\begin{tabular}{|c|c|c|c|c|c|c|}
\hline 
\multicolumn{7}{|c|}{Data Overview}\tabularnewline
\hline 
 &  &  & $\sqrt{s_{NN}}$ [TeV]  & Norm $\sigma$  & No Points  & Ref.\tabularnewline
\hline 
\hline 
ATLAS  & Run I  & $W^{\pm}$  & 5.02  & 2.7\%  & 10+10  & \cite{AtlasWpPb}\tabularnewline
\hline 
ATLAS  & Run I  & $Z$  & 5.02  & 2.7\%  & 14  & \cite{Aad:2015gta}\tabularnewline
\hline 
CMS  & Run I  & $W^{\pm}$  & 5.02  & 3.5\%  & 10+10  & \cite{Khachatryan:2015hha}\tabularnewline
\hline 
CMS  & Run I  & $Z$  & 5.02  & 3.5\%  & 12  & \cite{Khachatryan:2015pzs}\tabularnewline
\hline 
CMS  & Run II  & $W^{\pm}$  & 8.16  & 3.5\%  & 24+24  & \cite{Sirunyan:2019dox}\tabularnewline
\hline 
ALICE  & Run I  & $W^{\pm}$  & 5.02  & 2.0\%  & 2+2  & \cite{Alice:2016wka,Senosi:2015omk}\tabularnewline
\hline 
LHCb  & Run I  & $Z$  & 5.02  & 2.0\%  & 2  & \cite{Aaij:2014pvu}\tabularnewline
\hline 
\end{tabular}
\caption{
The overview of the LHC $W^\pm /Z$ data sets including the $p$Pb system
with per nucleon center-of-mass enery $\sqrt{s_{NN}}$, experimental normalization
uncertainty of the data,
number of data points, and references. 
}
\label{tab:data}
\end{table}
}  %
\def\tabNorm{
\begin{table}[t]
\begin{tabular}{|c|c|c|c|c|c|c|c|c|}
\hline 
\multicolumn{9}{|c|}{Normalization Shifts}\tabularnewline
\hline 
 & \multicolumn{3}{c|}{ATLAS Run I} & \multicolumn{3}{c|}{CMS Run I} & \multicolumn{2}{c|}{CMS Run II}\tabularnewline
\hline 
 & $W^{-}$  & $W^{+}$  & $Z$  & $W^{-}$  & $W^{+}$  & $Z$  & $W^{-}$  & $W^{+}$\tabularnewline
\hline 
Set ID & 6211  & 6213  & 6215  & 6231  & 6233  & 6235  & 6232  & 6234\tabularnewline
\hline 
\hline 
{\bf Norm0}  & \multicolumn{3}{c|}{---} & \multicolumn{3}{c|}{---} & \multicolumn{2}{c|}{---}\tabularnewline
\hline 
{\bf Norm2}  & \multicolumn{3}{c|}{0.963} & \multicolumn{3}{c|}{0.949} & \multicolumn{2}{c|}{---}\tabularnewline
\hline 
{\bf Norm3}  & \multicolumn{3}{c|}{0.955} & \multicolumn{3}{c|}{0.937} & \multicolumn{2}{c|}{0.960}\tabularnewline
\hline 
\end{tabular}
\caption{
The LHC \wz\ data sets and the corresponding ID's are listed.
The normalization factors (where appropriate) are applied to the data sets.
The normalization shifts for the {nCTEQ15WZ} set are the same as for the {Norm3} fit.}
\label{tab:norm}
\end{table}
}  %
\def\tabChi{
\begin{table*}[htb]
\begin{tabular}{|c|c|c|c|c|c|c|c|c|c|c|c|c|c|c|c|c|c|c|}
\hline 
\multicolumn{19}{|c|}{$\chidof$ for Selected Experiments \& Processes}  \tabularnewline
\hline
 & \multicolumn{3}{c|}{ATLAS Run I} & \multicolumn{3}{c|}{CMS Run I} & \multicolumn{2}{c|}{CMS Run II} & \multicolumn{2}{c|}{ALICE} & LHCb &
 & DIS & DY & Pion & LHC & LHC & {\bf~Total~}    \tabularnewline
\cline{1-12} \cline{2-12} \cline{3-12} \cline{4-12} \cline{5-12} \cline{6-12} \cline{7-12} \cline{8-12} \cline{9-12} \cline{10-12} \cline{11-12} \cline{12-12} \cline{12-12} 
 & $W^{-}$ & $W^{+}$ & $Z$ & $W^{-}$ & $W^{+}$ & $Z$ & $W^{-}$ & $W^{+}$ & $W^{-}$ & $W^{+}$ & $Z$ & &  &  &  &  &  {\small Norm} $\chi^{2}$  & \tabularnewline
\hline \hline
nCTEQ15 & 1.38 & 0.71 & 2.88 & 6.13 & 6.38 & 0.05 & 9.65 & 13.20 & 2.30 & 1.46 & 0.70 & & 0.91 & 0.73 & 0.25 & 6.20 & ---  & {\bf 1.66} \tabularnewline
\hline\hline
Norm0 & 0.94 & 0.26 & 2.71 & 3.89 & 2.25 & 0.03 & 1.07 & 1.51 & 0.53 & 0.02 & 0.71 & & 0.92 & 0.95 & 0.83 &  1.47 & ---   &  {\bf 1.03} \tabularnewline
\hline 
Norm2 & 0.78 & 0.37 & 1.90 & 1.60 & 1.02 & 0.21 & 1.25 & 1.59 & 0.59 & 0.02 & 0.68  & & 0.93 & 0.87 & 0.61 & 1.15 & 13.2 &  {\bf 0.98} \tabularnewline
\hline 
Norm3 & 0.83 & 0.46 & 1.78 & 1.41 & 1.23 & 0.31 & 0.74 & 0.84 & 0.75 & 0.08 & 0.62 & & 0.90 & 0.77 & 0.39 & 0.91 & 22.9  &  {\bf 0.91} \tabularnewline
\hline \hline
nCTEQ15WZ & 0.84 & 0.46 & 1.79 & 1.43 & 1.22 & 0.31 & 0.72 & 0.80 & 0.80 & 0.11 & 0.62 & & 0.90 & 0.78 & 0.38 & 0.90 & 22.9  &  {\bf 0.91} \tabularnewline
\hline \hline
\end{tabular}
\caption{
We present the $\chi^2/N_{dof}$ for the individual data sets,
the individual processes \{DIS, DY, Pion, LHC\}, and the total.
We also show the total $\chi^2$ contribution from the LHC normalization penalty.
Note that the Pion $\chi^2/N_{dof}$  is shown for comparison, but  this data is only
included in the {nCTEQ15WZ} fit.
\label{tab:Chi}
}
\end{table*}
}  %
\section{Introduction}
\begin{figure*}[htb]
\begin{center}
\subfloat[$\sqrt{s}=5.02$ TeV]{
\includegraphics[width=0.32\textwidth]{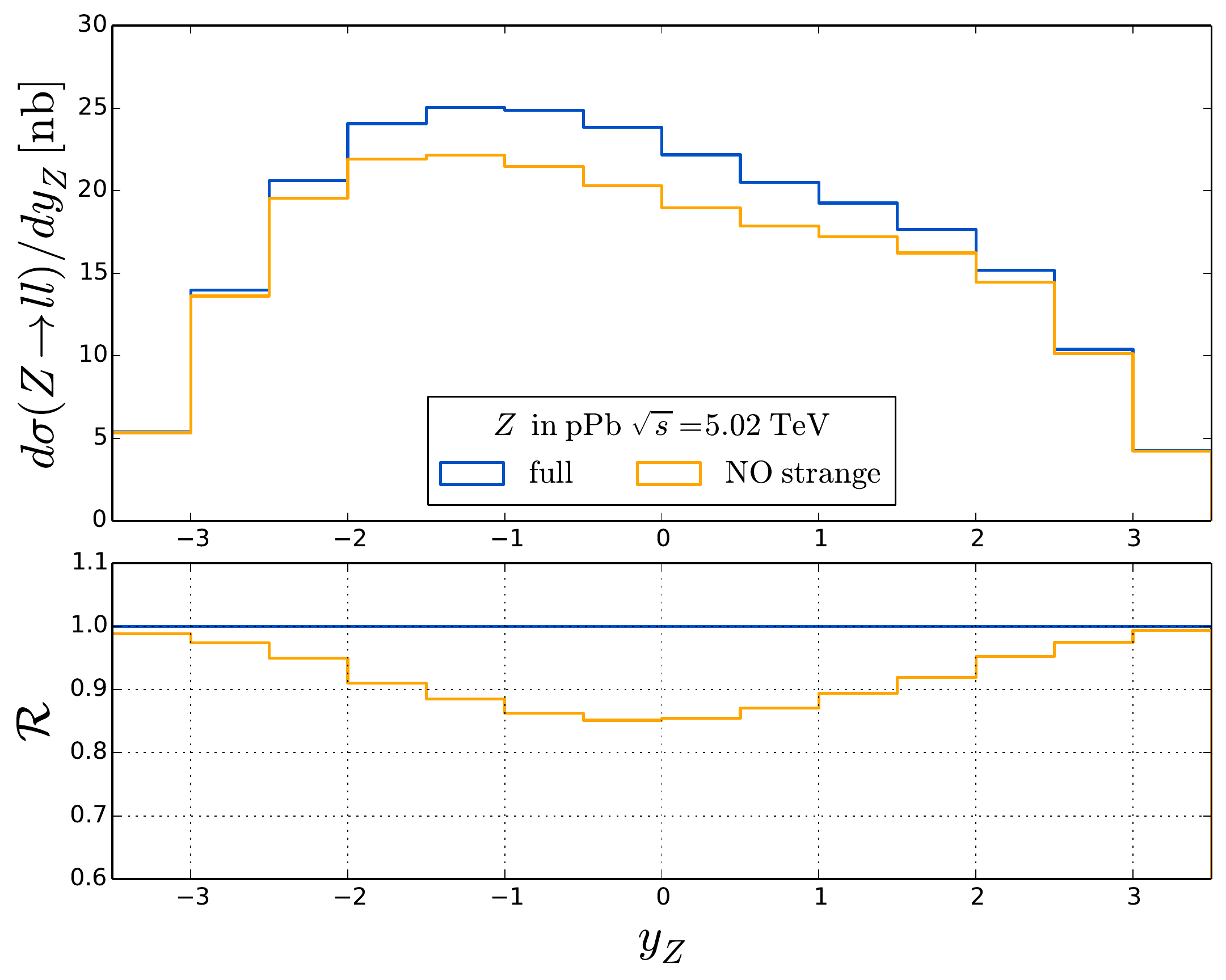}
\hfil
\includegraphics[width=0.32\textwidth]{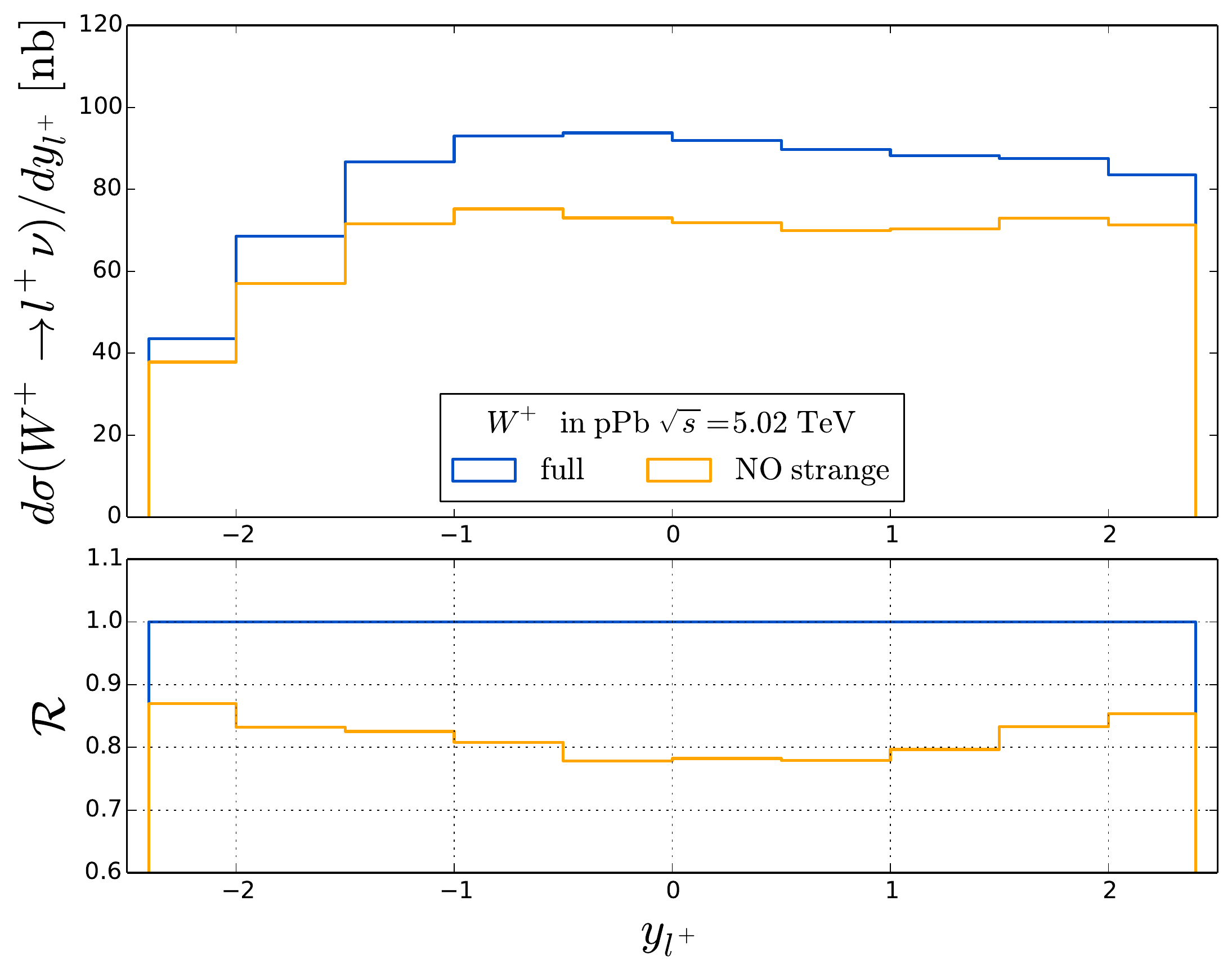}
\hfil
\includegraphics[width=0.32\textwidth]{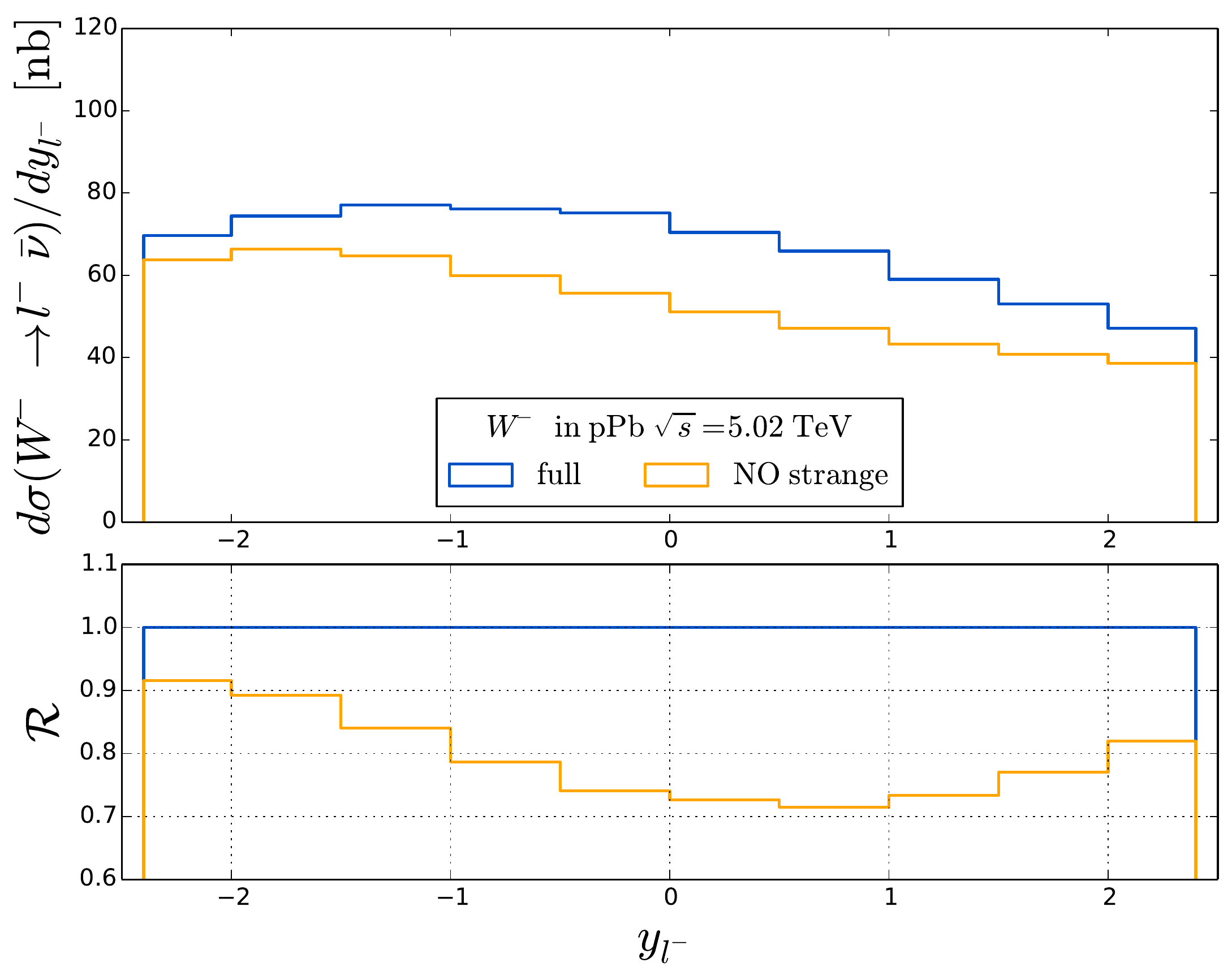}
\label{fig:STRcontr5TeV}}
\\
\subfloat[$\sqrt{s}=8.16$ TeV]{
\includegraphics[width=0.32\textwidth]{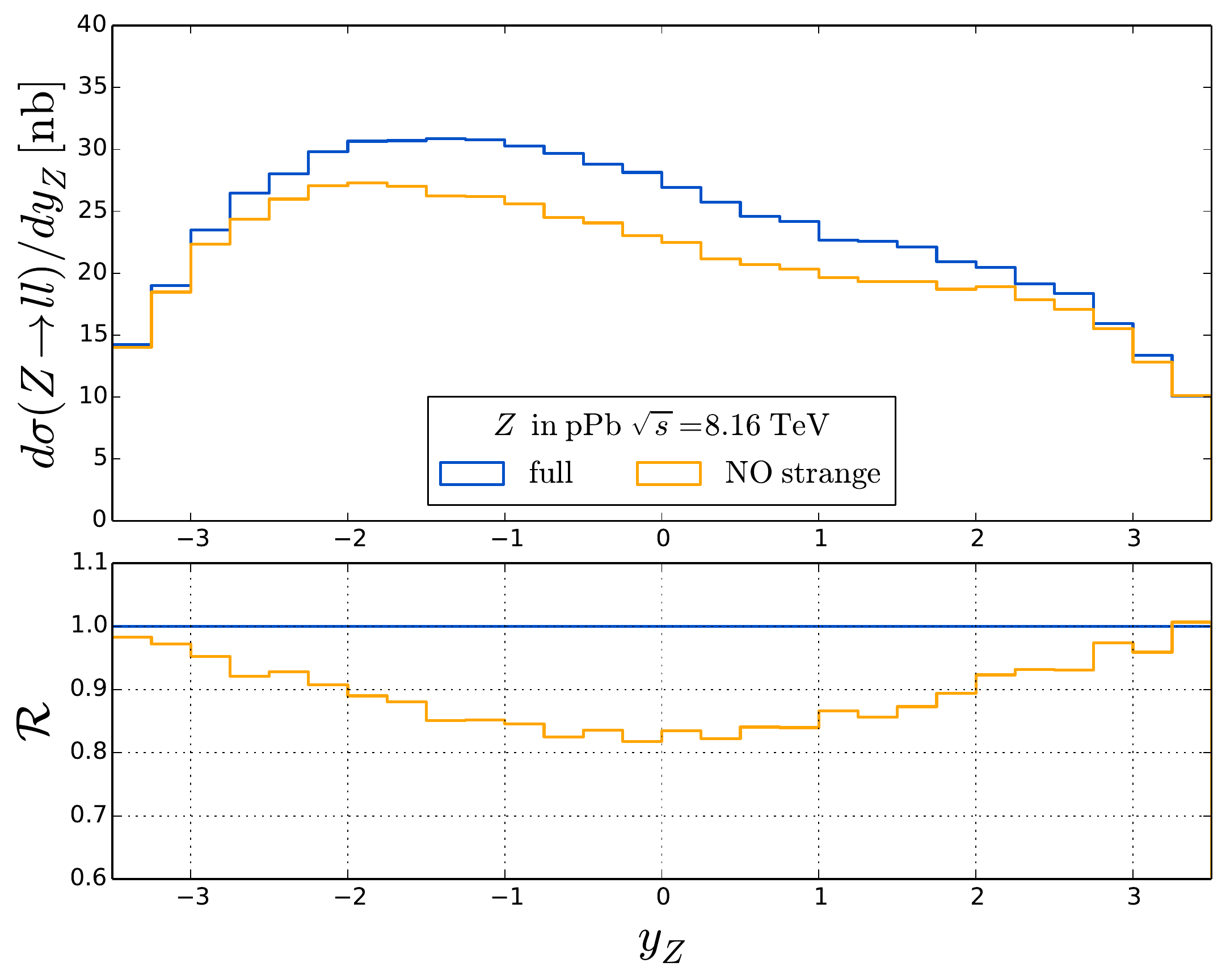}
\hfil
\includegraphics[width=0.32\textwidth]{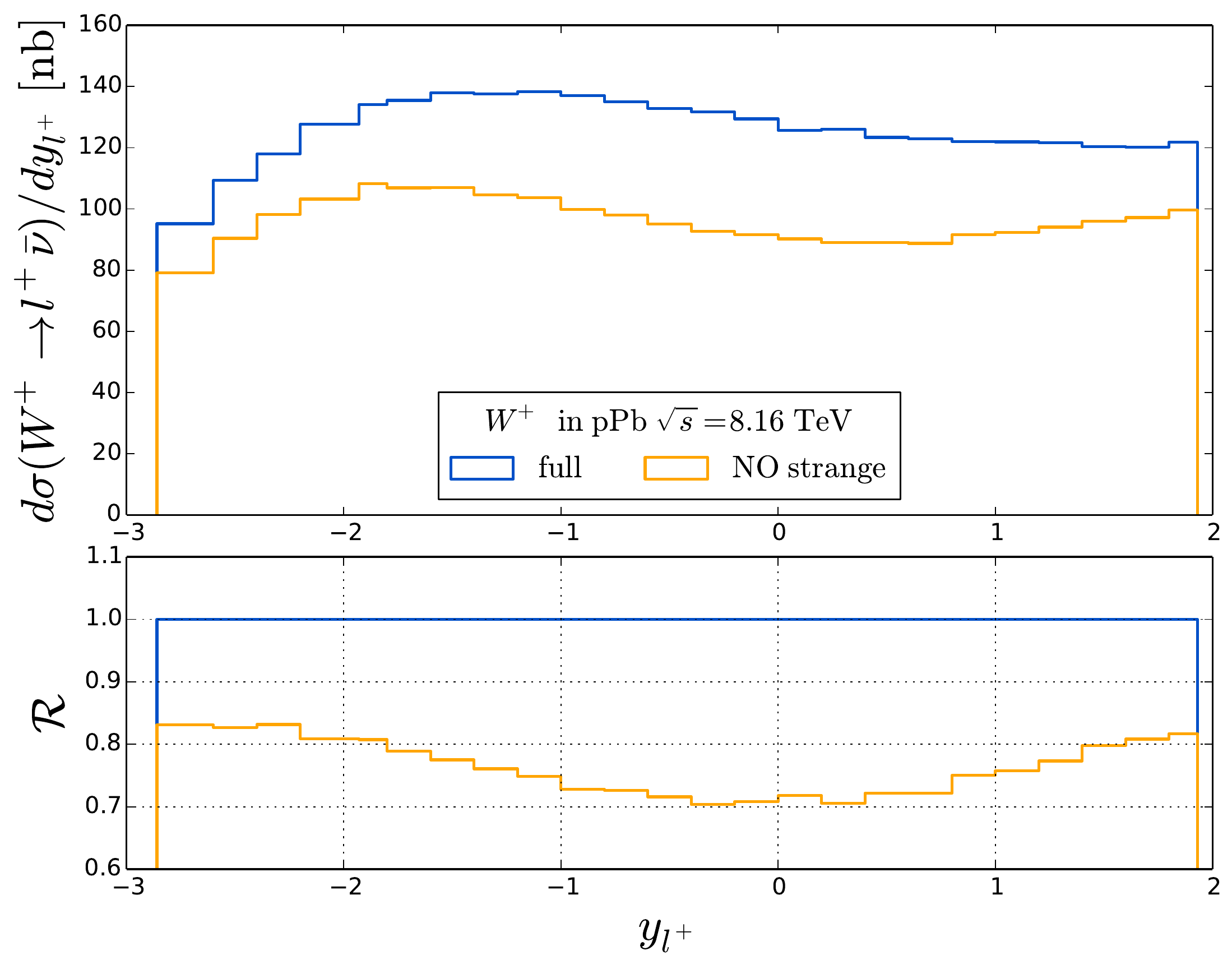}
\hfil
\includegraphics[width=0.32\textwidth]{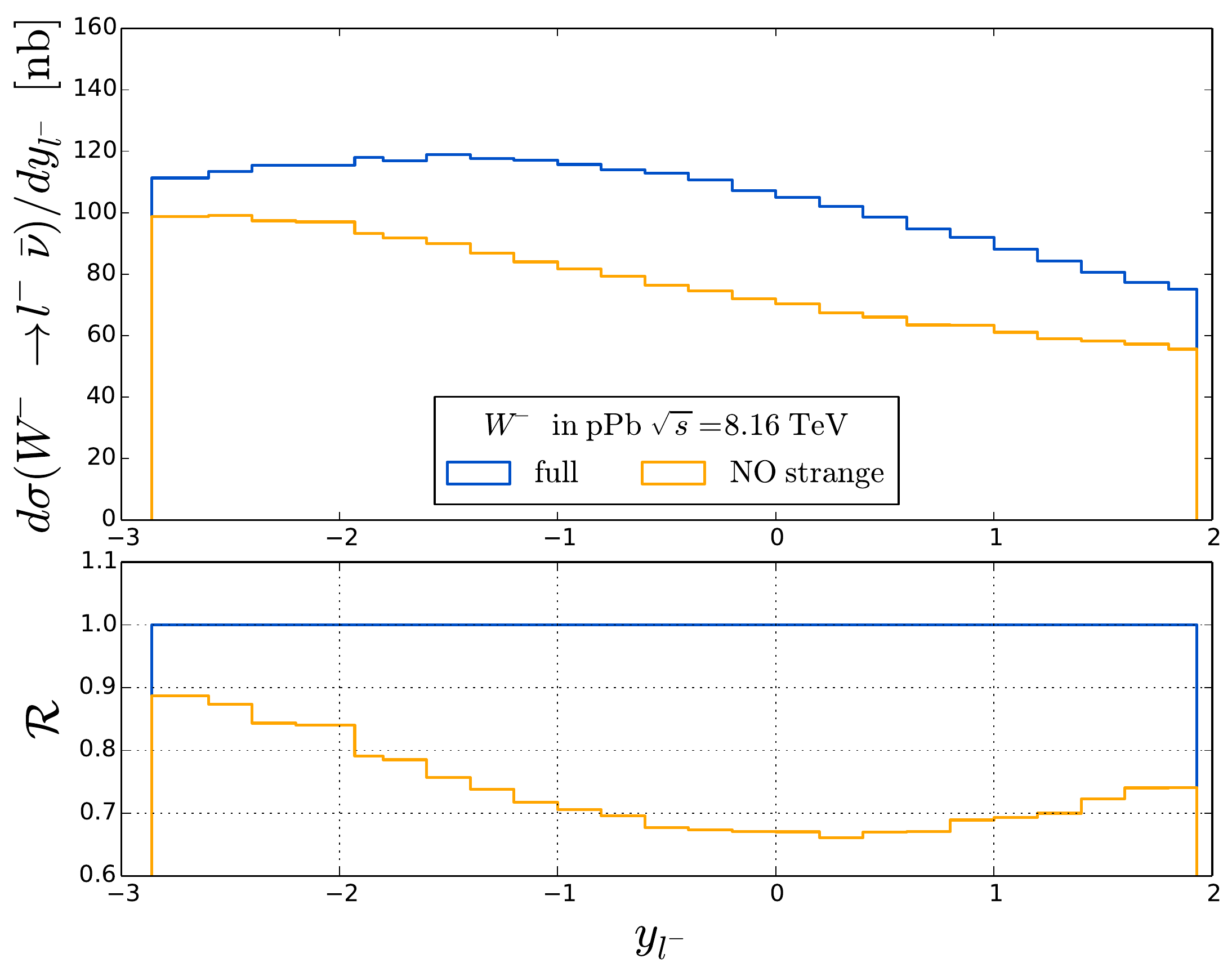}
\label{fig:STRcontr8TeV}}
\caption{Contribution of strange initiated channels to $W^{\pm}$ and $Z$ boson
production for proton-lead ($p$Pb) at the LHC. The blue lines represent total cross-sections,
the yellow lines are cross-sections with the strange initiated channels subtracted.
The lower panels show the ratio compared to the total cross-section.}
\label{fig:STRcontr}
\end{center}
\end{figure*}

Parton distribution functions (PDFs) are key elements required to generate
concrete predictions for processes with hadronic initial states in
the context of QCD factorization theorems.
The success of this theoretical framework has been extensively demonstrated
in fixed-target and collider experiments ({\it e.g.}, at the TeVatron, SLAC, HERA, RHIC, LHC),
and will be essential for making predictions for future facilities (EIC, LHeC, FCC). 
Despite the above achievements, there is yet much to learn about
the hadronic structure and the detailed composition of the
PDFs~\cite{Hou:2019efy,Ball:2017nwa,Kovarik:2015cma,Eskola:2016oht,AbdulKhalek:2019mzd,AbdulKhalek:2020yuc,Ethier:2020way,Khalek:2018mdn,Gao:2017yyd,Kovarik:2019xvh,Alekhin:2017olj,Hou:2019efy,Nadolsky:2008zw,Sato:2019yez,Harland-Lang:2014zoa,Thorne:2019mpt,Ball:2009mk,Lin:2017snn,Lin:2020rut}.

Although the up and down PDF flavors are generally well-determined across much of the
partonic $x$ range, there is significant uncertainty in the strange component, $s(x)$.
The strange PDF is especially challenging because, in many processes, it is difficult
to separate it from the larger down component. 
However, as we push to higher precision and energies,
an accurate determination of the strange PDF is, apart from its intrinsic
fundamental importance, essential not only for LHC measurements, but for
a wide variety of processes~\cite{Alekhin:2017olj,Hou:2019efy,Nadolsky:2008zw,Sato:2019yez,Harland-Lang:2014zoa,Thorne:2019mpt,Ball:2009mk,Lin:2017snn,Lin:2020rut}.
For example, the knowledge of the nuclear strange distribution in heavy nuclei
is crucial for providing a reliable baseline for hard probes of the quark gluon
plasma (QGP) which is characterized by enhanced production of
strangeness~\cite{Rafelski:1982pu,Deak:2017dgs,Gale:2020xlg}.
Additionally, small~$x$ nuclear PDFs are essential for computing the
composition of air showers from  ultra-high energy cosmic rays~\cite{Bhattacharya:2016jce,Reno:2019jtr,Bai:2020ukz,Zenaiev:2019ktw}.

The recent results from the LHC for $W/Z$ boson production in $pp$ collisions
predict a large strange to light-sea
ratio~\cite{Aad:2012sb,Aad:2014xca,Aaboud:2016btc,ATLAS:2019ext,Chatrchyan:2013uja,Sirunyan:2018hde}.
This is a rather surprising result as it differs from earlier determinations
based on analyses of neutrino deep inelastic scattering (DIS) data from NuTeV
and CCFR experiments~\cite{Tzanov:2005kr,Mason:2007zz,Goncharov:2001qe}
or charged kaon production data from HERMES~\cite{Airapetian:2008qf}.
See Ref.~\cite{Kusina:2012vh} for more details on the earlier determinations of the 
strange distribution,
and  Ref.~\cite{Cooper-Sarkar:2018ufj} for a study of the
compatibility of the ATLAS and CMS results using the xFitter
    framework~\cite{Alekhin:2014irh}.

It is not easy to directly compare the $pp$ LHC results with the fixed target
experiments, as the earlier measurements were generally done using nuclear targets
(typically Fe or Pb).
Additional complications arise from the fact that there is a controversy
about the proper nuclear correction factors for the charged current (CC)
and neutral current (NC) DIS measurements~\cite{Kovarik:2010uv,Schienbein:2009kk,Kalantarians:2017mkj,Paukkunen:2013grz,Paukkunen:2010hb}.
As a result, the choice of heavy target neutrino DIS data sets
varies widely among not just the many nuclear PDF (nPDF) determinations,
but also for the proton PDF fits~\cite{Hou:2019efy,Ball:2017nwa}.
Moreover, in the proton case the nuclear corrections are applied in
different ways.

Conversely, it was already demonstrated that the $W/Z$ LHC data can provide
some important information on the strange 
and gluon nPDFs~\cite{Kusina:2016fxy,Eskola:2016oht,AbdulKhalek:2020yuc}.
To demonstrate the  impact of the heavy ion \wz\  data on the strange PDF,
in Fig.~\ref{fig:STRcontr} we display the contribution of the strange-initiated
process  as a function of rapidity. We observe the
strange component can be as much as 20\% to 30\% of the total.
For this reason, we concentrate in the following on the constraints for the
nuclear strange and gluon distributions given by the $W$ and $Z$ data from
proton-lead ($p$Pb) collisions at the LHC.
This process is an ideal QCD ``laboratory'' 
as it  is sensitive to i)~the heavy flavor components $\{s,c,...\}$,
ii)~the nuclear corrections, and iii)~the underlying ``base'' proton PDFs. 
Such an analysis provides an independent perspective on the subject
and can help disentangle the flavor separation and nuclear modifications.

In the current investigation, we will study the
production of $W$ and $Z$ bosons in 
proton--lead ($p$Pb) collisions at the LHC;
this involves similar considerations as the $pp$ case,
but also brings in the nuclear corrections. 
We will be focusing, in particular, on the strange and gluon distributions
to see how these are modified when the LHC measurements are included. 
In Sec.~\ref{sec:fit} we review the various data sets used in our analysis
along with the separate fits extracted.
In Sec.~\ref{sec:results} we present the quality of the fits and comparisons
of data with the theory, and demonstrate the impact on the resulting PDFs. 
In Sec.~\ref{sec:other} we compare our final PDF fit with other results from the literature.
In Sec.~\ref{sec:conclusion} we recap the key outcomes of this study.
\section{Fits to Experimental Data \label{sec:fit}}

\subsection{The nCTEQ++ Framework}
The  nCTEQ project
extends the  proton PDF global fitting effort by fully including the nuclear
dimension.\footnote{For details, see \href{http://www.ncteq.org}{\texttt{www.ncteq.org}}
  which is hosted at \href{http://www.HepForge.org}{HepForge.org}.} 
Previous to the nCTEQ effort, nuclear data was ``corrected'' to isoscalar data 
and added to the proton PDF fit {\it without} any uncertainties~\cite{Olness:2003wz}.
In contrast, the nCTEQ framework allows full communication between the nuclear data 
and the proton data; this enables us to investigate if observed tensions between
data sets could potentially be  attributed to the nuclear corrections.

The details of the nCTEQ15 nPDFs are presented in Ref.~\cite{Kovarik:2015cma}.
The present analysis is performed in a new C++ code base  (\ncteqpp{})
which  enabled us to easily interface to external programs such as
HOPPET~\cite{Salam:2008qg},
APPLgrid~\cite{Carli:2010rw}, and MCFM~\cite{Campbell:2015qma}.
The nCTEQ15 fit has been reproduced in this new nCTEQ++ framework.

For the current set of fits, we use the same 16 free parameters as for the nCTEQ15 set,
and additionally open up three parameters for the strange PDF, for a total of 19 parameters.
Recall that for the nCTEQ15 set, the strange PDF was constrained 
by the relation $s=\bar{s}=(\kappa/2)(\bar{u}+\bar{d})$
at the initial scale $Q_0=1.3$ GeV
so that it had the same form as the sea quarks. 

Our PDFs are parameterized as
\begin{equation}
    x f_i^{p/A}(x,Q_0) = 
    c_0 x^{c_1} (1-x)^{c_2} e^{c_3 x} (1+ e^{c_4} x)^{c_5}
    \quad ,
\end{equation}
and the nuclear $A$ dependence is encoded in the coefficients as 
\begin{equation}
    c_k \longrightarrow c_k(A) \equiv c_{k,0} + c_{k,1} (1-A^{-c_{k,2}})
    \quad ,
\end{equation}
where $k=\{1, ... ,5\}$.

The 16 free parameters used for the nCTEQ15 set
model the $x$-dependence of the
$\{g, u_v, d_v, \bar{d}+\bar{u} \}$ PDF combinations, and
we do not vary the $\bar{d}/\bar{u}$ parameters;
see Ref.~\cite{Kovarik:2015cma}   for details.
To this, we now add three strange PDF parameters: $\{c_{0,1}^{s+\bar{s}},c_{1,1}^{s+\bar{s}},c_{2,1}^{s+\bar{s}} \}$;
these parameters describe, correspondingly, the overall normalization,
  the low-$x$ exponent and the large~$x$ exponent of the strange distribution.

\subsection{Experimental Data Sets}
\label{sec:data}
\tabData
\tabNorm
\tabChi

In this analysis we use the deep inelastic scattering (DIS),  
Drell-Yan (DY) lepton pair production, and RHIC pion  data employed in our earlier nCTEQ15
analysis~\cite{Kovarik:2015cma}.
Additionally, we use 
$W$ and $Z$ inclusive data from proton-lead collisions at the LHC.
Specifically, we include the following data sets:
ALICE $W^{\pm}$ boson production~\cite{Alice:2016wka,Senosi:2015omk},
ATLAS $Z$ boson production~\cite{Aad:2015gta},
ATLAS $W^{\pm}$ boson production~\cite{AtlasWpPb},
CMS $Z$ boson production~\cite{Khachatryan:2015pzs},
CMS $W^{\pm}$ boson production~\cite{Khachatryan:2015hha},
CMS Run II $W^{\pm}$ boson production~\cite{Sirunyan:2019dox},
and 
LHCb $Z$ boson production~\cite{Aaij:2014pvu}.
The data sets are outlined in Table~\ref{tab:data}.
All the theory calculations are performed at the next-to-leading order (NLO)
of QCD. In particular, the calculations for $W$ and $Z$ boson production
have been performed using the MCFM-6.8 program~\cite{Campbell:2011bn} interfaced
with APPLgrid~\cite{Carli:2010rw} in order to speed up the computations.

\subsection{The PDF Fits}

We now use the \ncteqpp{} framework to include the LHC \wz{} $p$Pb data
and extend the nCTEQ15 fit.
Comparing the LHC $p$Pb data to the nCTEQ15 results,
we find that these data generally lie above  the theory
predictions~\cite{Kusina:2016fxy};
hence, if we allow for a normalization uncertainty,
this additional freedom can significantly improve the fit.
The experiments have an associated luminosity uncertainty
(\textit{cf.,} Table~\ref{tab:norm}), and we will use
this as a gauge as we shift the data normalizations.
It is reasonable to tie the normalizations for $\{W^{\pm}, Z\}$
data from individual experiments (e.g, CMS Run~I)
to a single normalization factor as these uncertainties are fully correlated. 

Previous studies implied a close connection between
the normalization of the \wz\ data and the extracted strange PDF~\cite{Kusina:2016fxy}.
To systematically investigate the effect of the normalization in detail,
we will use a series of fits outlined in Table~\ref{tab:Chi}
and summarized below:
\begin{description}[leftmargin=1.5cm,labelindent=0.5cm]

\item[{\bf nCTEQ15}] This is the original set of nuclear PDFs
  as computed in Ref.~\cite{Kovarik:2015cma}.

\item[{\bf Norm0}]
  We include the LHC $p$Pb data, but we do not
  allow for any floating normalization of the LHC data.

\item[{\bf Norm2}]
  We include the LHC $p$Pb data,
  and  allow for 2 normalization factors;
  one for the ATLAS Run~I data, 
  one for CMS Run~I;
  we do not renormalize CMS Run~II data in this fit. 
  
\item[{\bf Norm3}]
  We include the LHC $p$Pb data,
  and we allow for 3 normalization factors;
  one for the ATLAS Run~I data, 
  one for the CMS Run~I data, and
  a separate one for the CMS Run~II data.
  
\item[{\bf nCTEQ15WZ}]
  This is the same as Norm3, but we also include the
  RHIC inclusive pion data directly in the fit.
  This is discussed in Sec.~\ref{sec:other}.
  
\end{description}
All four of these new PDF fits are based on the DIS and DY data from the
nCTEQ15 analysis and the LHC data sets, as outlined in Sec.~\ref{sec:data}
and Table~\ref{tab:data}.

As with our nCTEQ15 study, we will present results both with and
without the inclusive pion data~\cite{Adler:2006wg,Abelev:2009hx}.
For the comparison of the the \wz\ normalizations fits \{{Norm0},
{Norm2}, {Norm3}\}, we will not include the pion data;
however, we do compute the pion $\chi^2$, as shown in Table~\ref{tab:Chi}, to
demonstrate the  compatibility.\footnote{%
  Note that nCTEQ15WZ extends nCTEQ15 by adding the LHC \wz\ data.
  In a similar manner, Norm3 extends the nCTEQ15np fit; however,
  we choose not to label this as nCTEQ15WZnp to avoid possible confusion. }
In Sec.~\ref{sec:other} we then present a separate fit, {nCTEQ15WZ}, which does
include the pion data.
As we will see the impact of the pion data is marginal.

\note{Normalization Factors}
Table~\ref{tab:norm} shows the determined normalization factors used
in each fit. All the normalization shifts are between $1 \sigma$ and $2 \sigma$
of the quoted normalization uncertainty, but all are systematically below unity;
the appropriate normalization penalties are included in the $\chi^2$ calculations.
The detailed prescription we use for fitting data normalizations is provided in
the Appendix~\ref{app:Norm}.
For the ALICE and LHCb sets, the current 
uncertainties provide sufficient flexibility that we
do not use an additional normalization factor for these data.

\section{Results and Discussion \label{sec:results}}

Having presented our series of fits, we now examine
i)~the quality of these fits as measured by the $\chi^2$ values,
ii)~the comparison of the data with our theory predictions,
and
iii)~the  impact on the underlying  PDFs.

\subsection{Quality of the fits}
\begin{figure*}[!p]
\begin{center}
\includegraphics[width=0.8\textwidth]{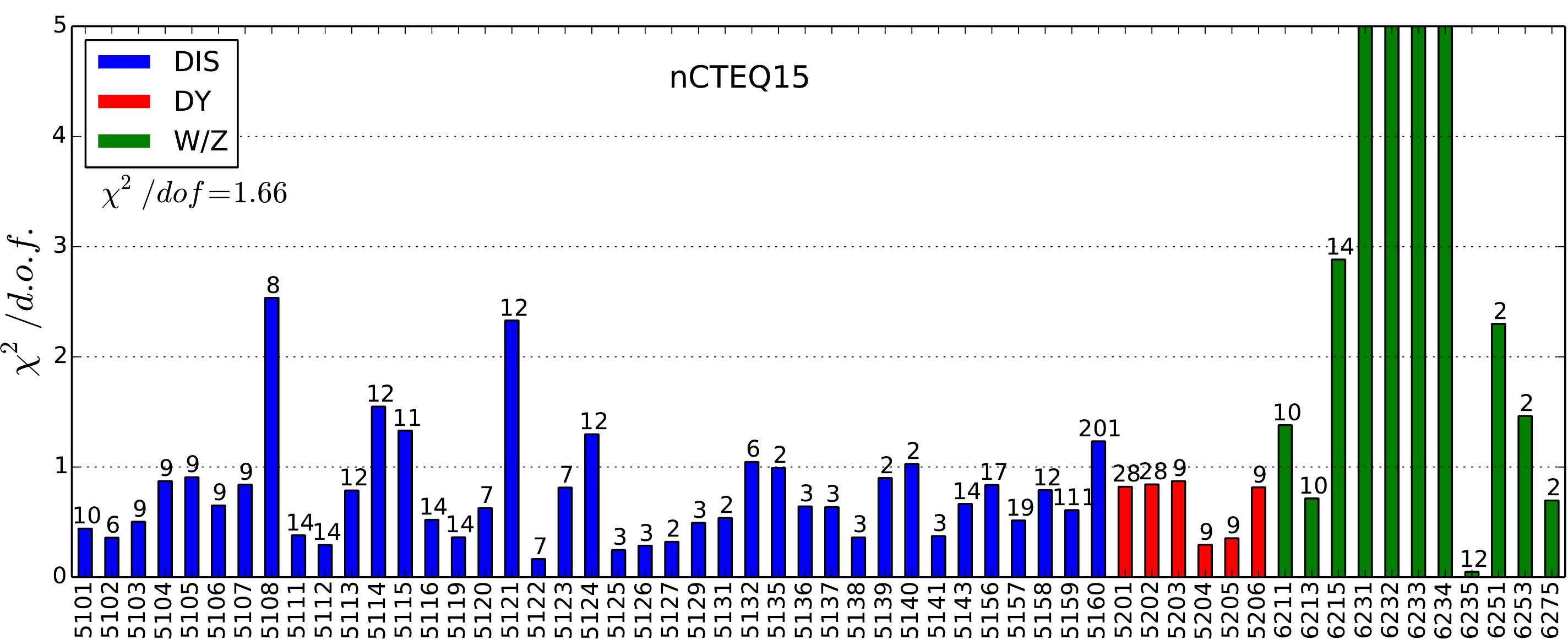}
\\
\includegraphics[width=0.8\textwidth]{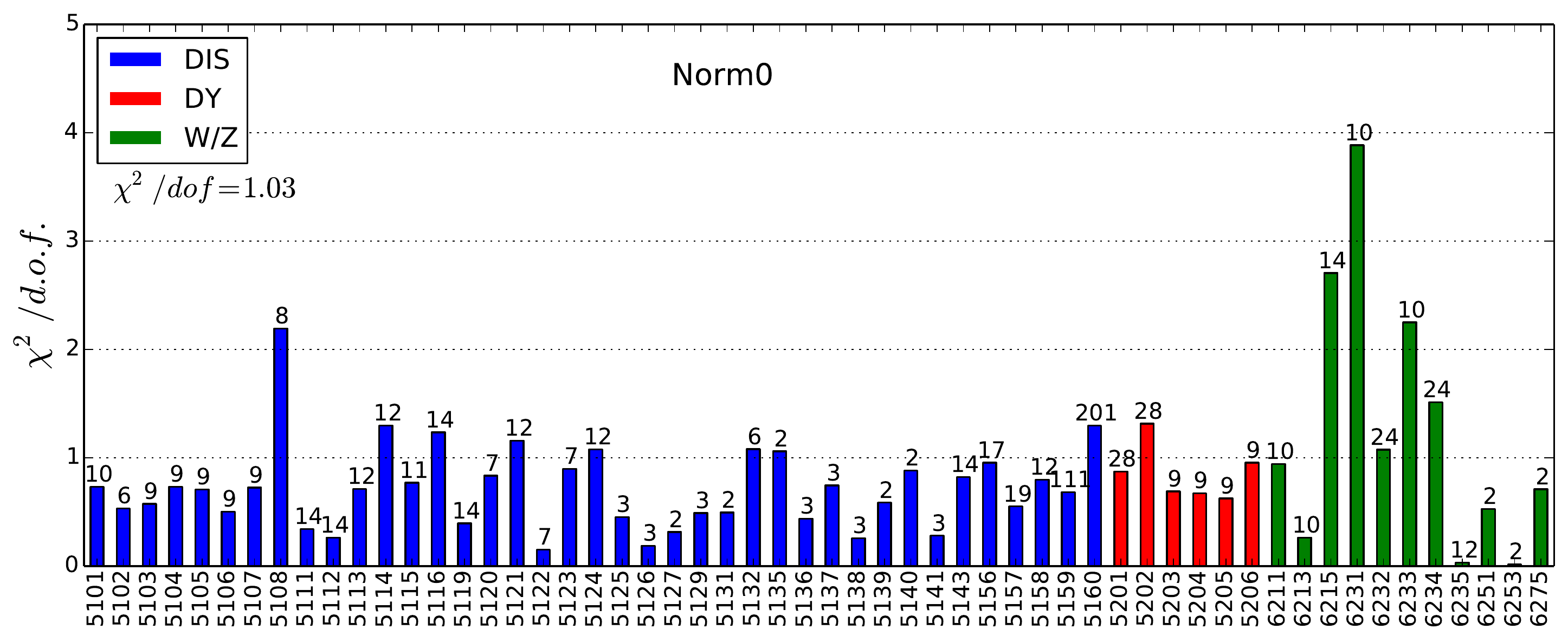}
\\
\includegraphics[width=0.8\textwidth]{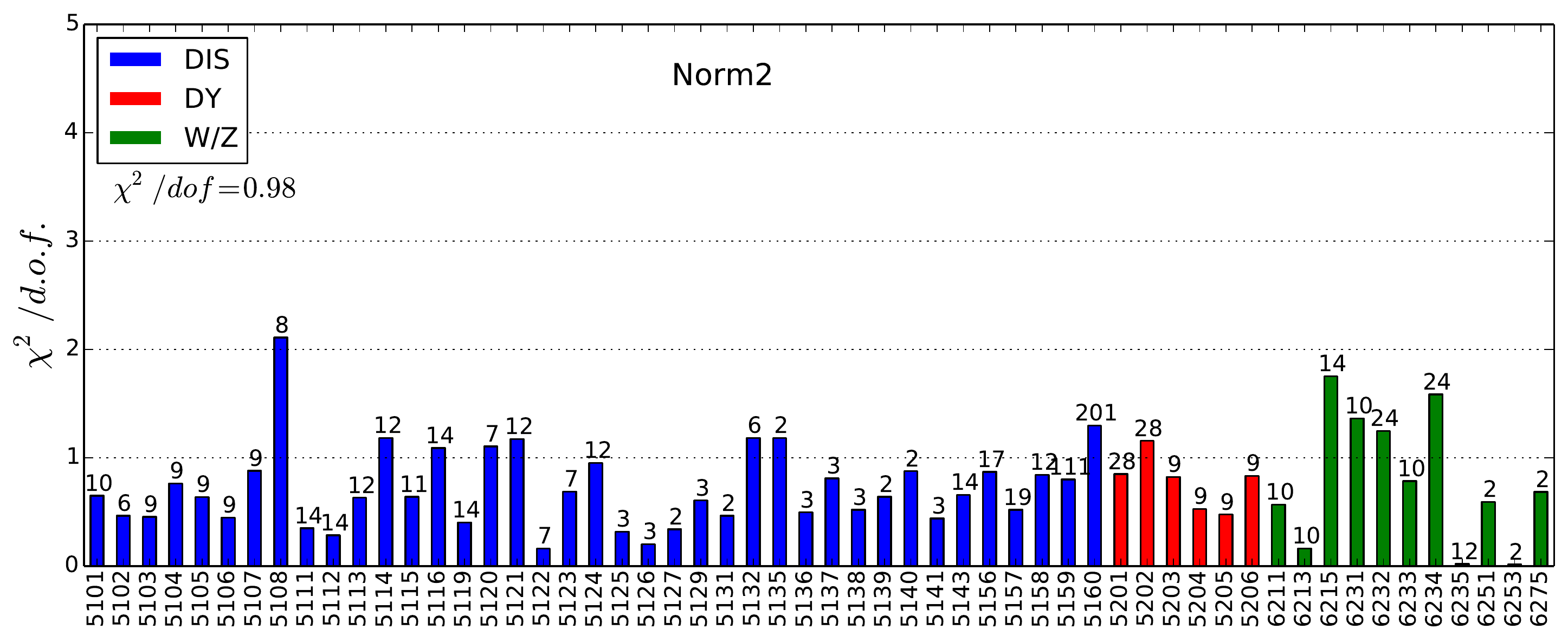}
\\
\includegraphics[width=0.8\textwidth]{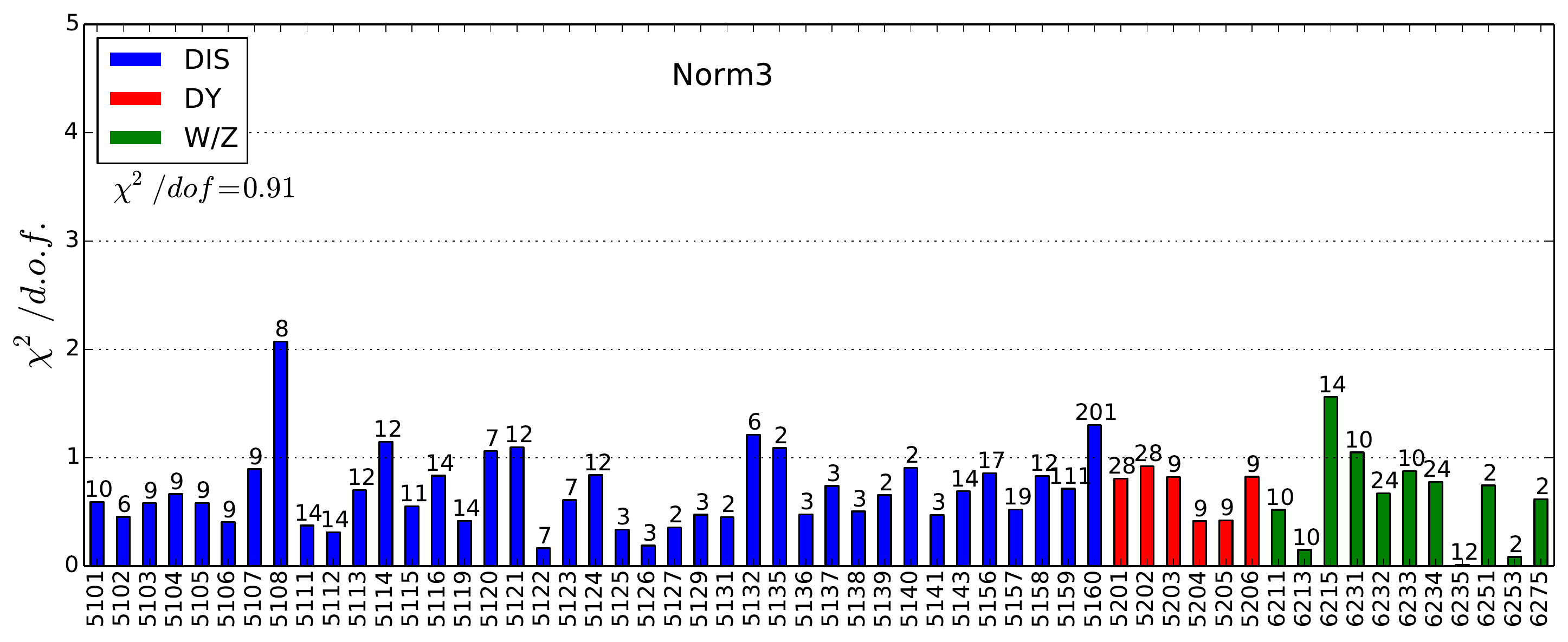}
\caption{The $\chidof$ of the individual experiments.
  The number of data points is indicated at the top of the bar,
  and the total $\chidof$ for the entire set is shown below the legend.
  For the nCTEQ15 set, four of the bars extend beyond the chart range;
  these are $\chi^2/N_{dof} = \{6.1, 9.7, 6.4, 13.2\}$
  for experiments $\{6231, 6232,6233,6234\}$.
}
\label{fig:chi2dofEXP}
\end{center}
\end{figure*}
\note{Overall quality of the fits}
In Tab.~\ref{tab:Chi} we present the $\chidof$ for selected data sets
as well as for each experiment type,%
\footnote{When we refer to $N_{dof}$ for the total $\chi^2$ we calculate
  it in the usual way as the difference between number of data points
  and number of free parameters ($N_{dof}=N_{data}-N_{par}$).
  However, when referring to $N_{dof}$ for individual experiments or data
  sets we set it to be equal to the number of data points ($N_{dof}=N_{data}$).}
and the contribution of the normalization penalty to the total $\chidof$.
We compute the normalization penalty as outlined in Appendix~\ref{app:Norm},
and this is included in the total. 

Examining the total  $\chidof$ of the fits, 
we see a broad range spanning from 1.66 for nCTEQ15 to values below 1.00,
and an even larger range of the  $\chidof$ for the  individual LHC data sets.

\note{Quality of individual data sets}
To provide more details regarding the source of the $\chi^2$
contributions, in Fig.~\ref{fig:chi2dofEXP} we display the $\chidof$
values for each
individual experiment which enters the fit.  The experiments are
identified by their 4-digit ID, and the number of data points is
indicated at the top of each bar.%
\footnote{The IDs of the specific non-LHC experiments can be found
  in Ref.~\cite{Kovarik:2015cma}.
  In general,  DIS data sets are 51xx,
  DY sets are 52xx,
  and \wz\ sets are 62xx.
}
Additionally, the bars are color-coded to indicate the type of observable
\{DIS, DY,W/Z\}.

The  $\chidof$ bar charts provide incisive information as to
which data sets are driving the fit.
We discuss each fit in turn. 

\note{nCTEQ15}
Starting with the nCTEQ15 set, we note that (except for a few outliers) 
the DIS and DY data is well described by these PDFs;\footnote{%
  We find the DIS experiments 5108 (Sn/D EMC-1988) and 5121 (Ca/D NMC-1995)
  to be outliers with  $\chidof>2$; this is consistent with other
  analyses~\cite{Eskola:2016oht,deFlorian:2011fp}.
  }
by comparison,  the  LHC $W^\pm /Z$ data (which was {\bf not} included in the original nCTEQ15 fit)
is not well described. 
As was detailed in Ref.~\cite{Kusina:2016fxy}, an important contribution  to this large~$\chi^2$  
comes from the small~$x$ region where the nuclear PDFs are poorly constrained.
The re-weighting analysis of Ref.~\cite{Kusina:2016fxy} demonstrated that we can  improve 
the fit by adjusting the small~$x$ behavior of the PDFs, but this alone will not bring all the data sets
into the range of $\chidof\sim 1$; something else is required.

\note{Norm0}
As a first step, this fit includes the LHC $W^\pm /Z$ data,
but does not include any floating normalization factors. 
This fit will tell us the extent to which we can adjust the PDFs to fit the
LHC data before we begin to adjust the normalization factors.
Examining Fig.~\ref{fig:chi2dofEXP}, we see the impact
of this fit on the DIS and DY data is generally small for many of the data sets,
but does result in noticeable improvement for a few of the sets
including 5115 (NMC Ca/D) and 5121 (NMC Li/D), for example.
However, it does significantly improve the LHC $W^\pm /Z$ fit reducing the
partial $\chi^2/N_{dof}$ of this data from 6.20 to 1.47 for the 120 LHC data points. 
Although this is a notable improvement, a number of the LHC data sets
still have  $\chidof$ values well above one.

\note{Norm2} In this fit, we allow two floating normalization factors
(one for ATLAS and one for CMS Run~I) which are allowed to vary in the fit.
The contribution of the normalization penalty  is included in the total $\chi^2$.

We see the impact of the floating normalization factors on the DIS and DY data is again small,
as was the case for the Norm0 fit.
But the Norm2 fit dramatically improves  the LHC $W^\pm /Z$ data reducing
$\chi^2/N_{dof}$  of this data to 1.15 as compared to 1.47 for the Norm0 fit.
While the Norm2 fit is a substantial improvement over the Norm0 results
and all LHC data sets have $\chi^2/N_{dof}<2$,  
there are still a few sets at the upper limit of this range. 

\note{Norm3}
Finally, we now perform a fit with three normalization factors:
one for ATLAS (Run~I), and one for each CMS Run~I and CMS Run~II. 

As before, the modifications to the DIS and DY sets are minimal,
but we do continue to see an improvement in the LHC sets;
namely the 
$\chi^2/N_{dof}$  of this data improves to 0.91 as compared to 1.15 for the Norm2 fit.

Comparing Norm3 with nCTEQ15 for the other data sets,
we see that the $\chi^2/N_{dof}$ for the DIS data is
essentially the same (0.91 vs. 0.90),
the DY increases slightly (0.73 vs. 0.77),
and the pion $\chi^2$  (computed a posteriori) increases
slightly as well (0.25 vs. 0.39);
these differences are relatively small compared to the significant 
improvement in the LHC data (6.20 vs. 0.90).
\subsection{Comparison of Data with Theory}
\begin{figure*}[tbp]
\begin{center}
\includegraphics[width=0.45\textwidth]{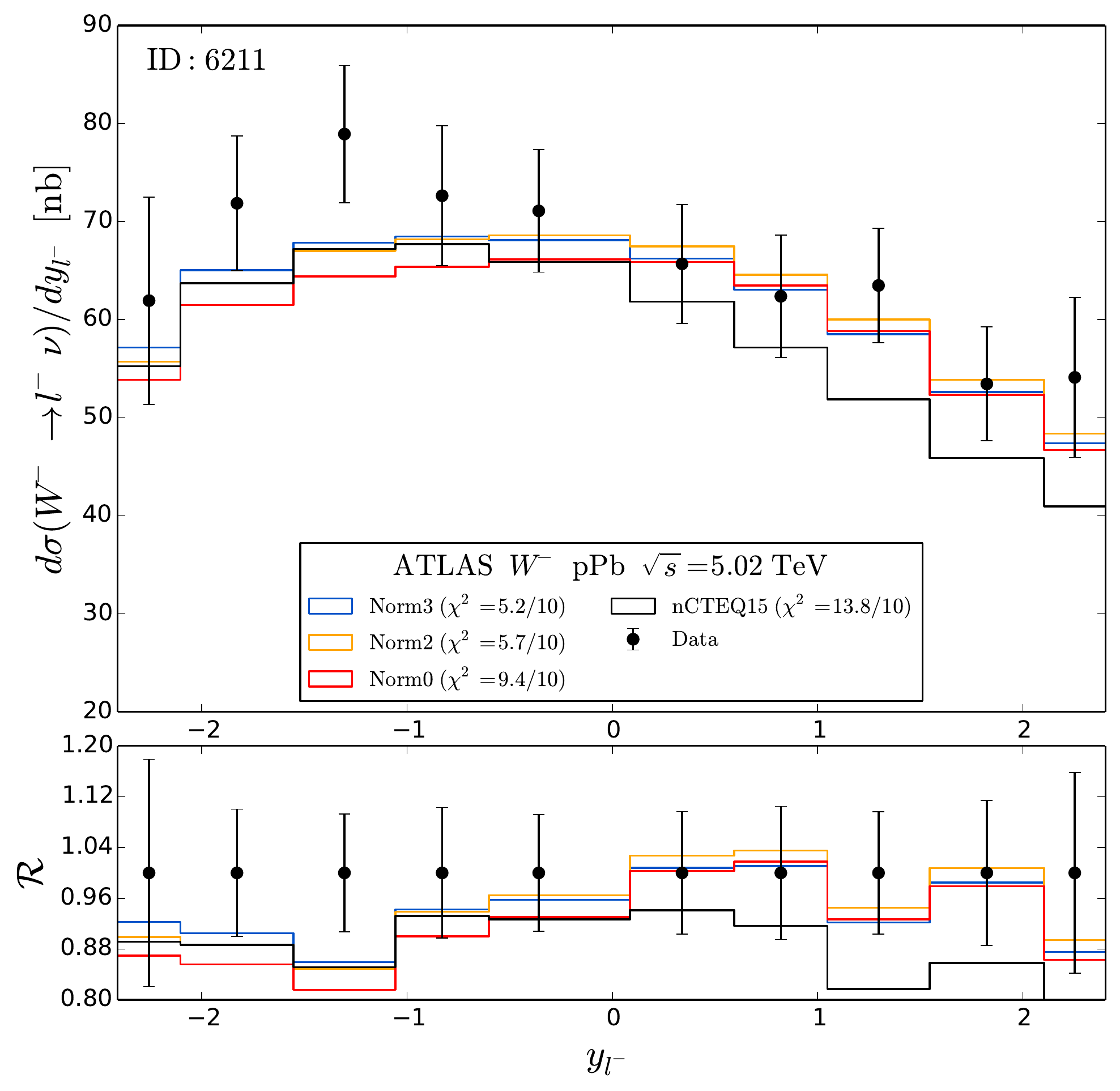}
\hfil
\includegraphics[width=0.45\textwidth]{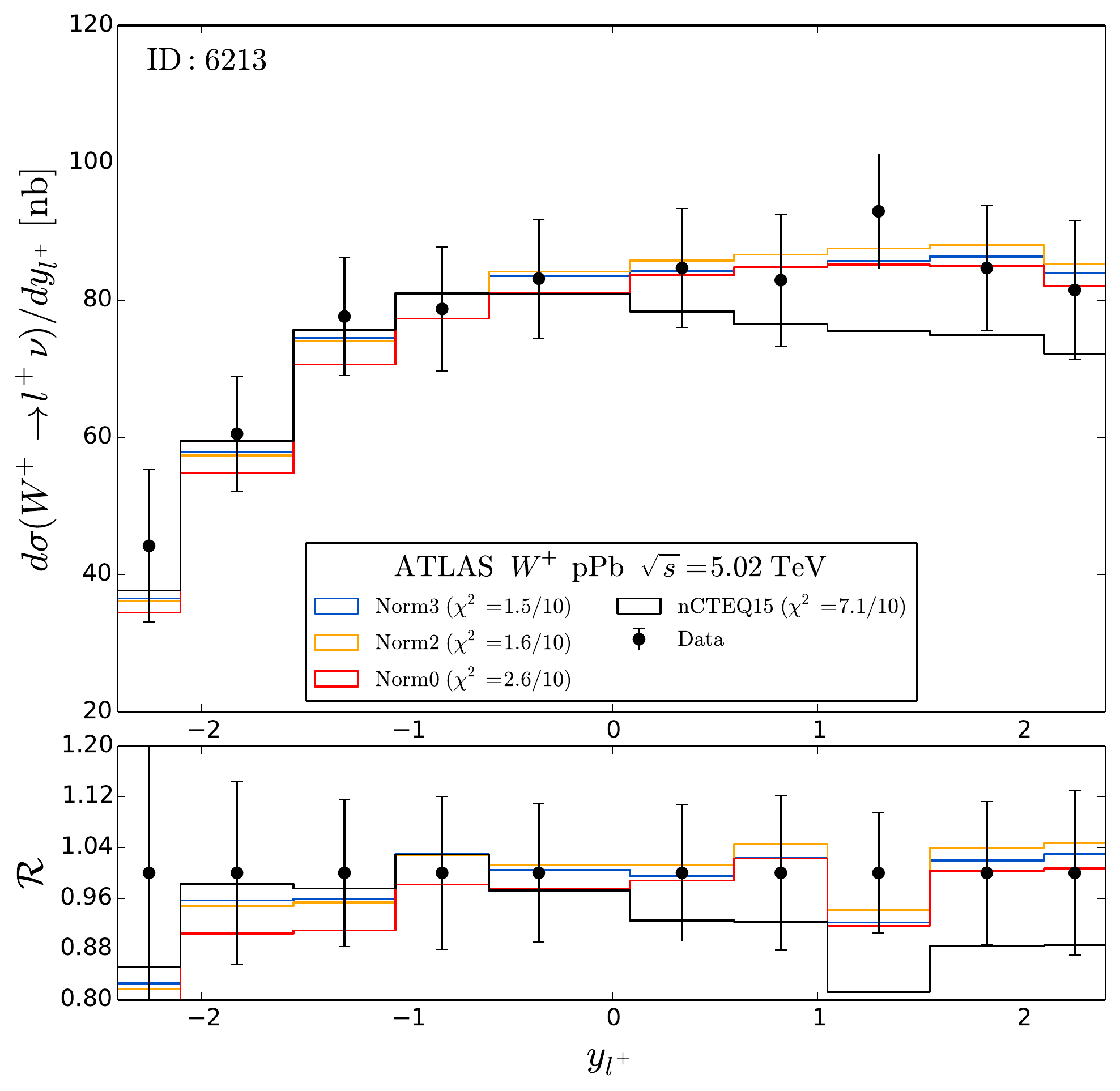}
\\
\includegraphics[width=0.45\textwidth]{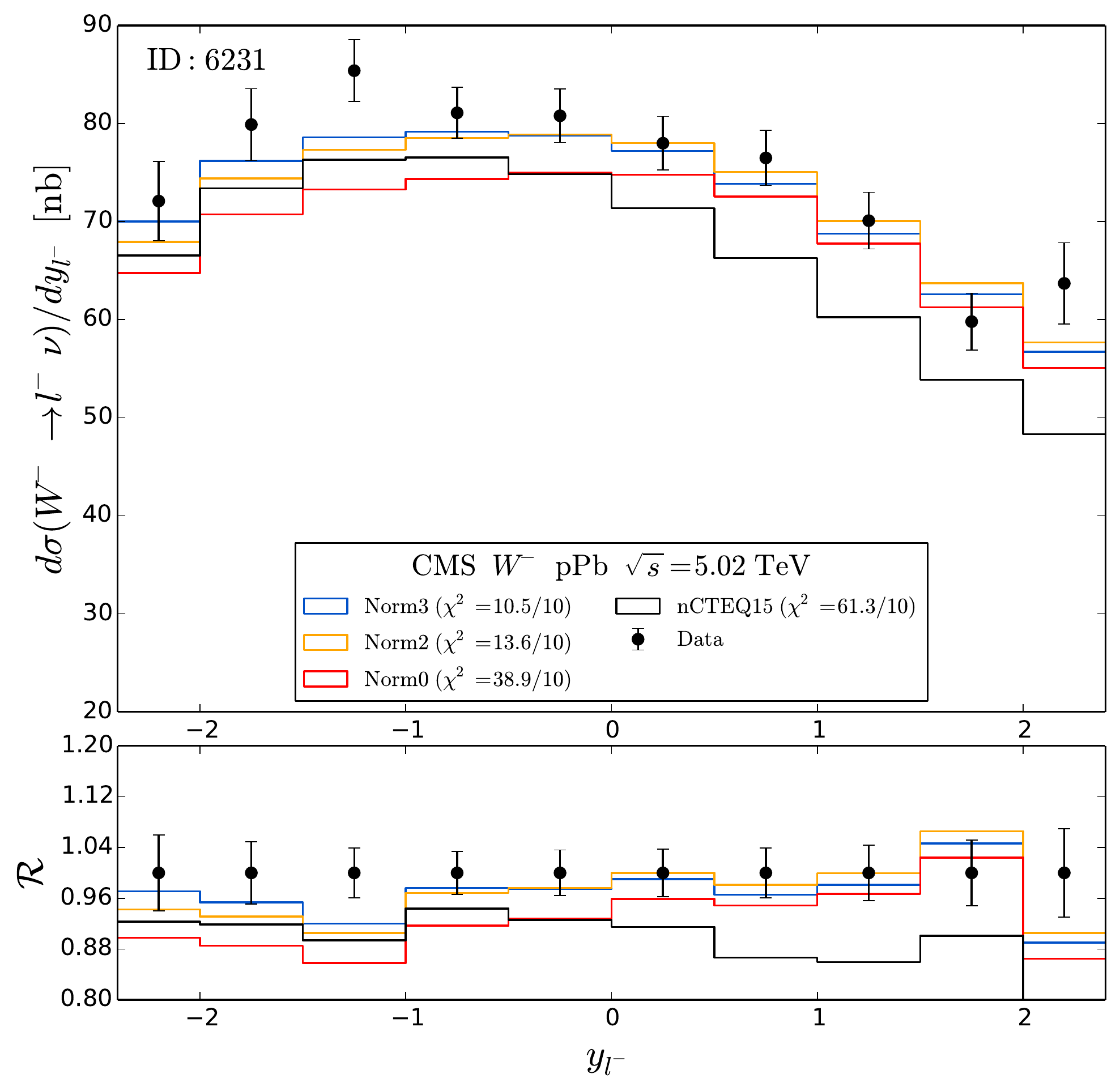}
\hfil
\includegraphics[width=0.45\textwidth]{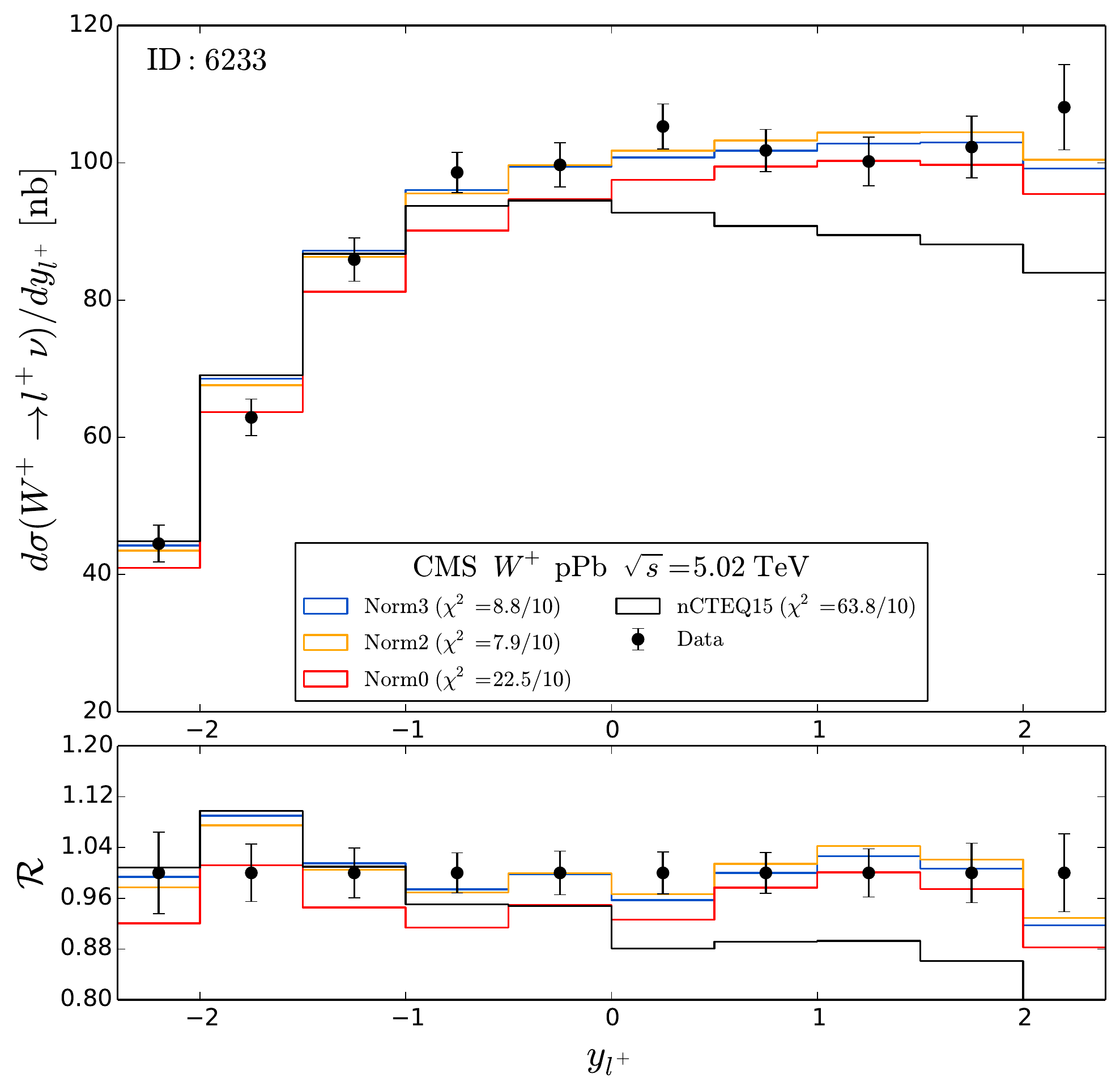}
\\
\includegraphics[width=0.45\textwidth]{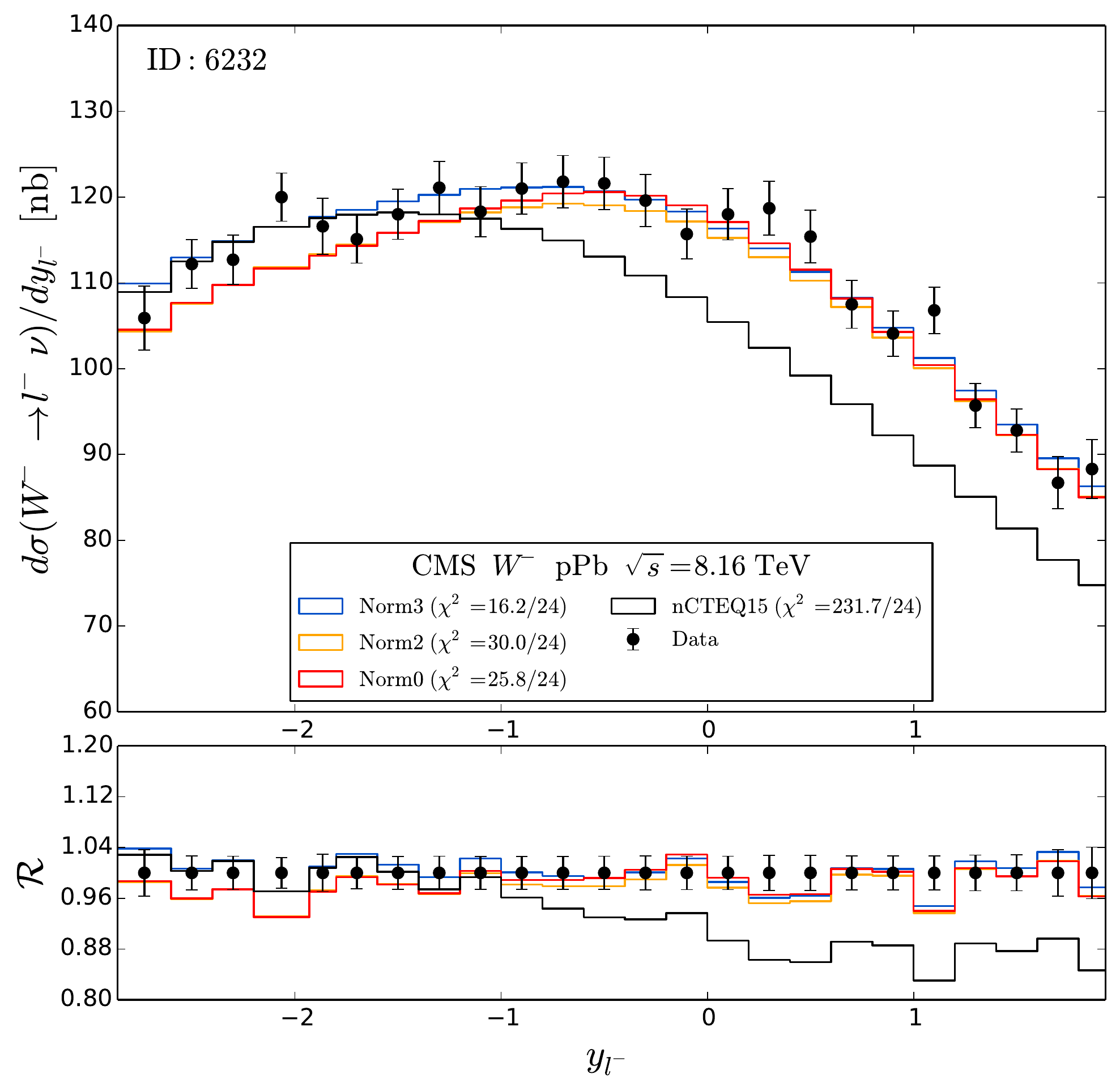}
\hfil
\includegraphics[width=0.45\textwidth]{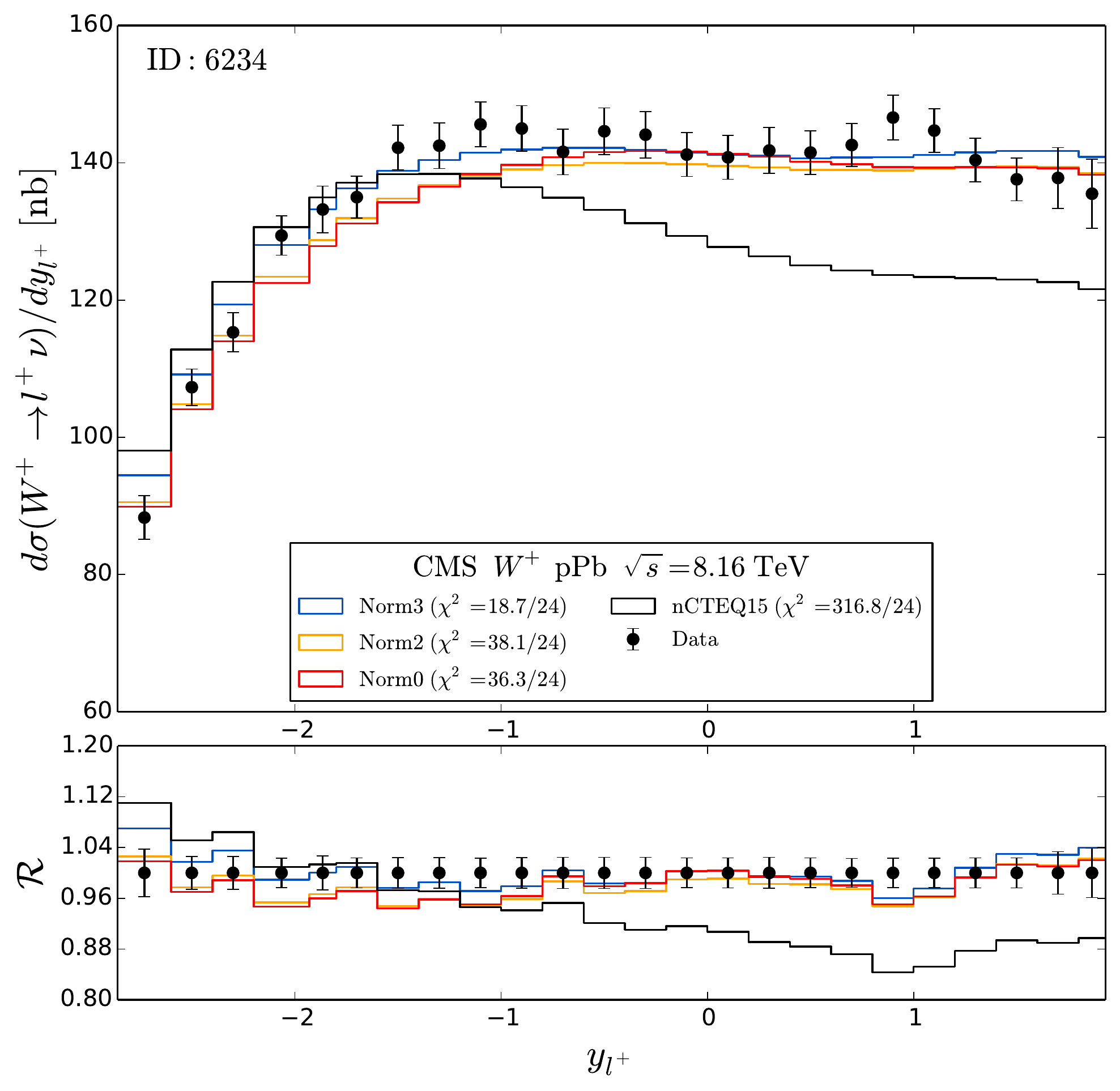}
\caption{Comparison of data with theory for ATLAS and CMS $W^\pm$ production.
  The normalization shifts are applied to the theory so we can compare all
  the results on a single plot; the data is unaltered. 
  For reference, 
  ATLAS Run~I $\{W^-,W^+\}=\{6211,6213\}$,
  CMS   Run~I $\{W^-,W^+\}=\{6231,6233\}$ and 
  CMS   Run~I $\{W^-,W^+\}=\{6232,6234\}$.}
\label{fig:dataOth_W}
\end{center}
\end{figure*}
\begin{figure*}[tb]
\begin{center}
\includegraphics[width=0.47\textwidth]{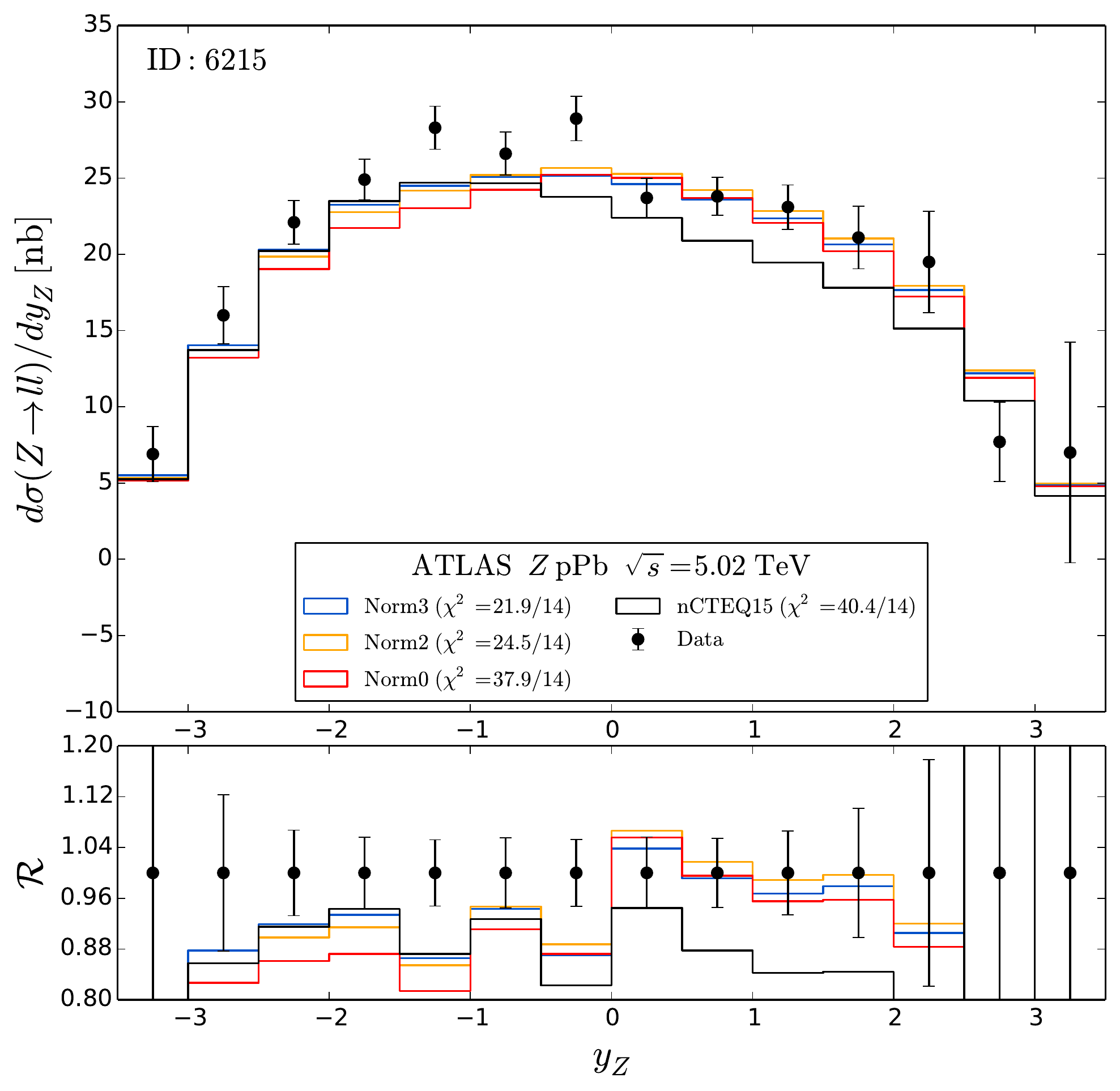}
\hfil
\includegraphics[width=0.47\textwidth]{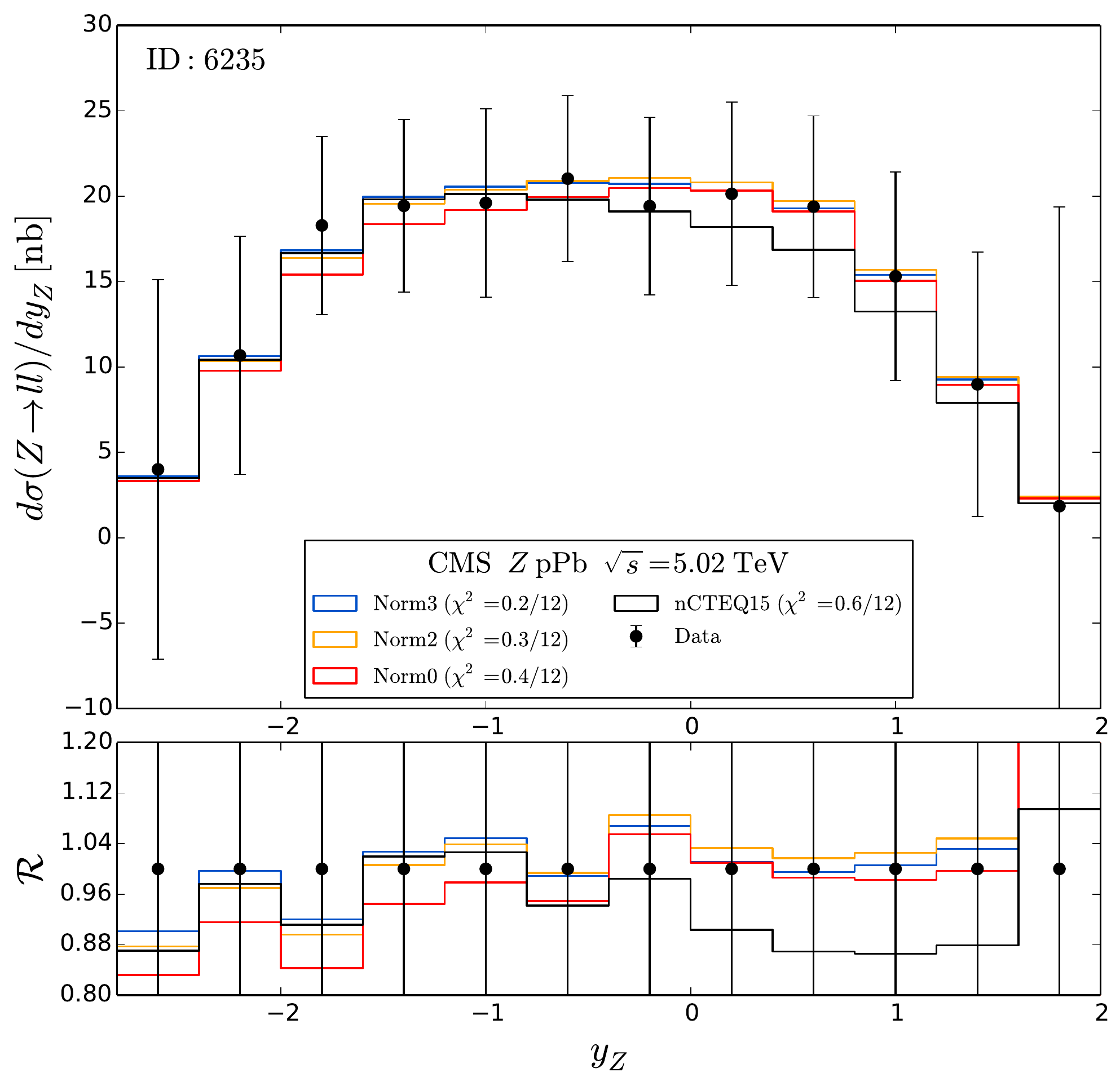}
\caption{
Comparison of data with theory for ATLAS and CMS $Z$ production.
  The normalization shifts are applied to the theory so we can compare all
  the results on a single plot; the data is unaltered. 
  For reference, 
  ATLAS Run~I $\{Z\}=\{6215\}$ and 
  CMS   Run~I $\{Z\}=\{6235\}$.}
\label{fig:dataOth_Z}
\end{center}
\end{figure*}
\begin{figure*}[tb]
\begin{center}
\includegraphics[width=0.47\textwidth]{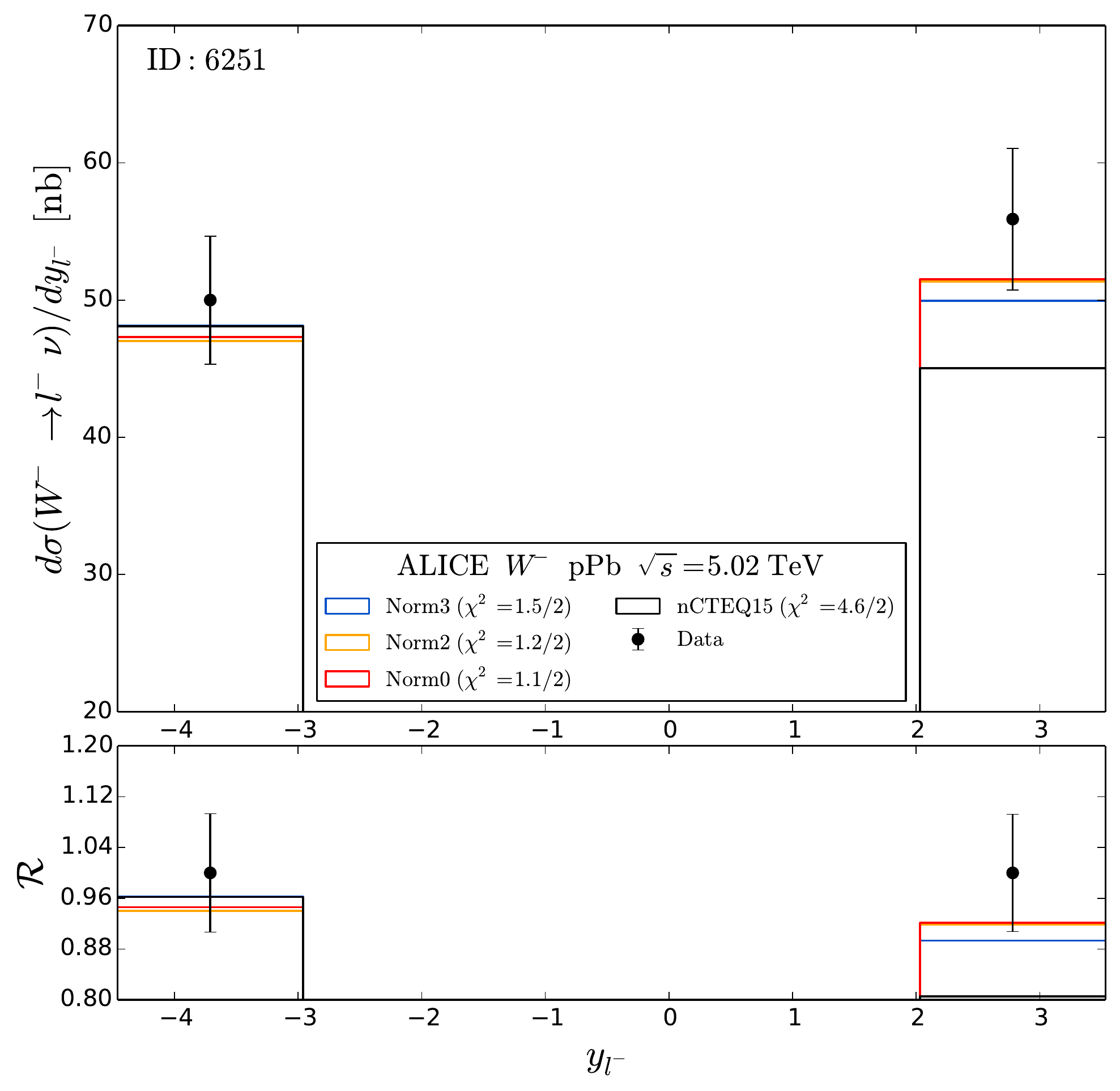}
\hfil
\includegraphics[width=0.47\textwidth]{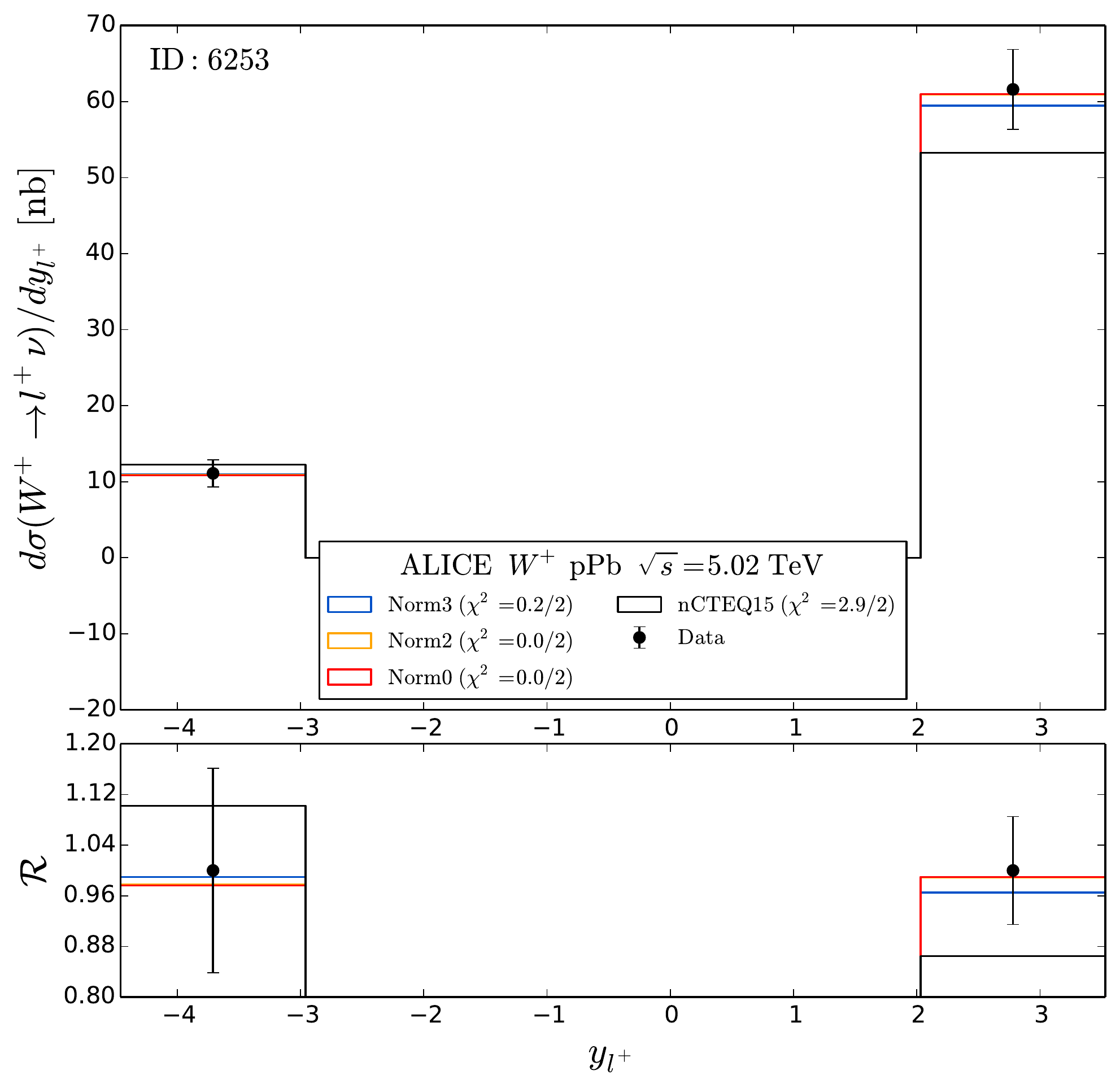}
\\
\caption{Comparison of data with theory for ALICE $W^\pm$ production.
  The normalization shifts are applied to the theory so we can compare all
  the results on a single plot; the data is unaltered. 
  For reference, ALICE Run~I $\{W^-,W^+\}=\{6251,6253\}$.}
\label{fig:dataOth_ALICE}
\end{center}
\end{figure*}
\begin{figure}[tb]
\begin{center}
\includegraphics[width=0.47\textwidth]{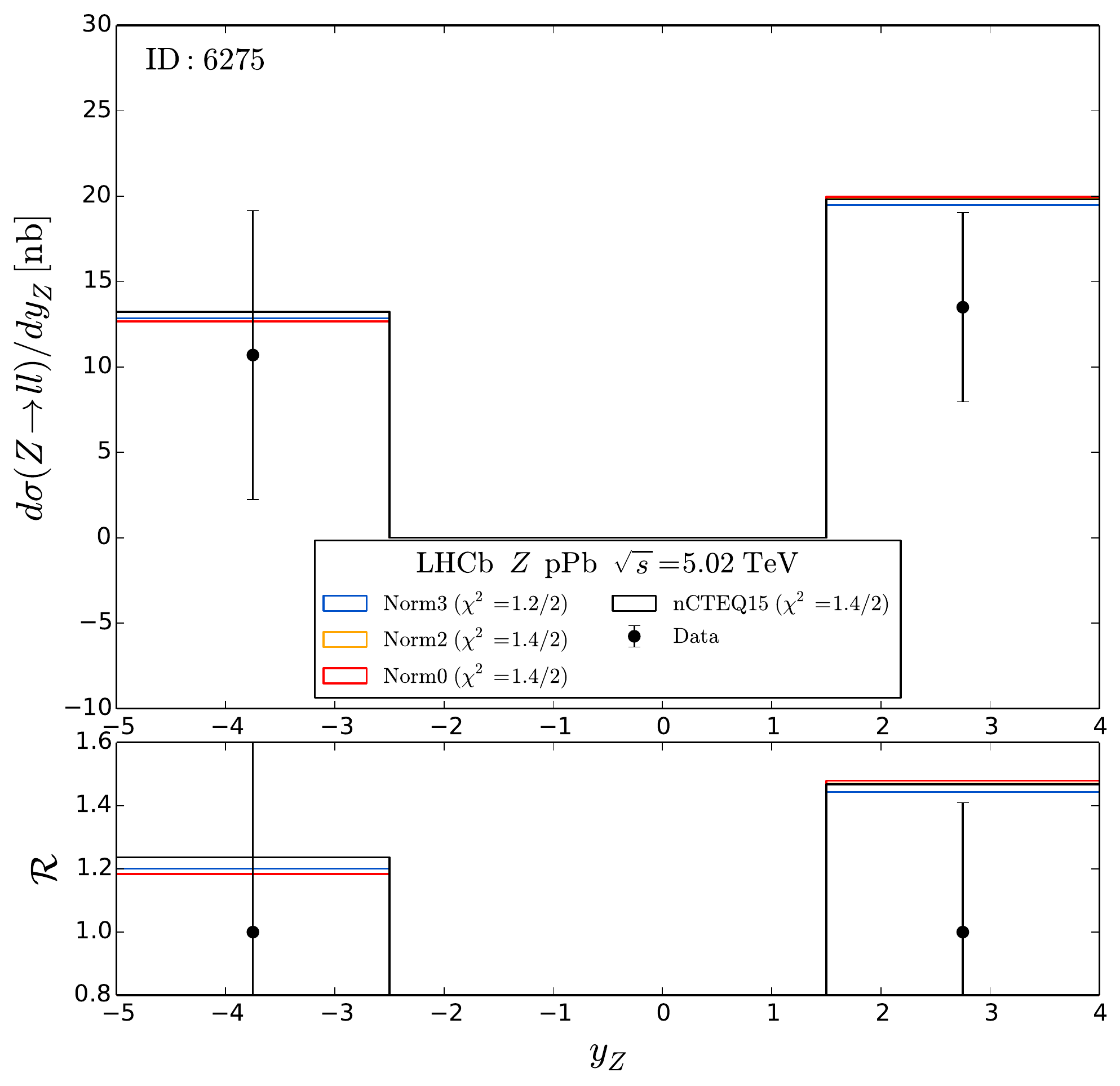}
\caption{Comparison of data with theory for LHCb $Z$ production (ID 6275).
  The normalization shifts are applied to the theory so we can compare all
  the results on a single plot; the data is unaltered.}
\label{fig:dataOth_LHCb}
\end{center}
\end{figure}

\vspace{10pt}

To obtain a more complete view of the fit quality, 
in
Figs.~\ref{fig:dataOth_W},
\ref{fig:dataOth_Z},
\ref{fig:dataOth_ALICE},
and~\ref{fig:dataOth_LHCb}
we display the comparison of the LHC data with theory predictions. 
The data points and errors are taken directly from the experimental measurements.
However, it is  important to note that we have shifted the theoretical predictions
by the appropriate normalization factors;
this allows us to present the fits with different normalizations on a single plot, and
provides a more accurate visual description of the quality of the fit. 

\subsubsection{Large~$x$ Region:}

Our first observation is that the experimental data
consistently lies above our theoretical predictions.
From \hbox{Table~\ref{tab:norm}}, recall that {\bf all} the fitted normalization factors
are less than one, indicating that the fit prefers a reduction of the data values, 
typically in the range of $\sim 5\%$;
because we have shifted the theory,
this is not as obvious in Figs.~\ref{fig:dataOth_W}--{\ref{fig:dataOth_LHCb}.

Even with the normalization shifts, we see the theory predictions still lie
well below the data for a number of sets.
This is most evident in the negative $y$ region for the Run~I $W^-$ data sets 
6211 (ATLAS $W^-$)
and
6231 (CMS $W^-$ Run~I),
and to a lesser extent 6215 (ATLAS $Z$).
Interestingly, the Run~II data generally show good agreement across the full $y$ range.

The negative rapidity region corresponds to the large~$x$ region of the lead PDF.
The large~$x$ region is already rather well constrained by the
fixed-target measurements, so there are limits
as to how much the new LHC data can shift the PDFs in this region.
Also note, that in the large~$x$ region  we are in the
``anti-shadowing region'' ($x\sim 0.1 $) where the nuclear corrections typically enhance the nuclear PDF
relative to the proton.
Thus, not including  the nuclear corrections in this region would increase the discrepancy. 

\subsubsection{Small $x$ Region:}

In the large rapidity (small $x$) region,
we generally find good agreement between our new fits and the data.
But, this is in striking contrast to the nCTEQ15 PDF which lies well below 
many of the data points at large~$y$;
this behavior is clearly evident, for example, in
6215 (\hbox{ATLAS}~$Z$),
and
6213, 6232, 6233, 6234 (CMS~$W^\pm$ Run~I and Run~II).
Clearly, the new LHC \wz\ data provides important new PDF constraints
in this kinematic region that were not available in the nCTEQ15 analysis.

As larger rapidity corresponds to smaller $x$ values,
this puts us in the ``shadowing region'' ($x\lesssim 0.1 $) where
the nuclear PDFs are generally expected to be suppressed relative to the proton.
If the nuclear shadowing correction were reduced in this region,
that would bring the theory closer in line with the data
{\it without} the need for large normalization factors. 
The precise value of the nuclear corrections is still an open question;
for example, 
Refs.~\cite{Schienbein:2009kk,Kovarik:2010uv,Owens:2007kp}
found that the shadowing correction for
the $\nu N$ charged-current neutrino DIS  was reduced as compared to the 
$\ell^\pm N$ neutral-current DIS.
If such an adjustment were applied to the LHC \wz\ data,
it would move the theory closer to the data and reduce the normalization factor.
Disentangling the nuclear effects from the underlying parton flavor components is intricate,
and a reanalysis of the neutrino DIS data is currently in progress~\cite{ncteqNeutrino}.

\subsection{The PDFs}
\begin{figure*}[tb]
\begin{center}
\includegraphics[width=\textwidth]{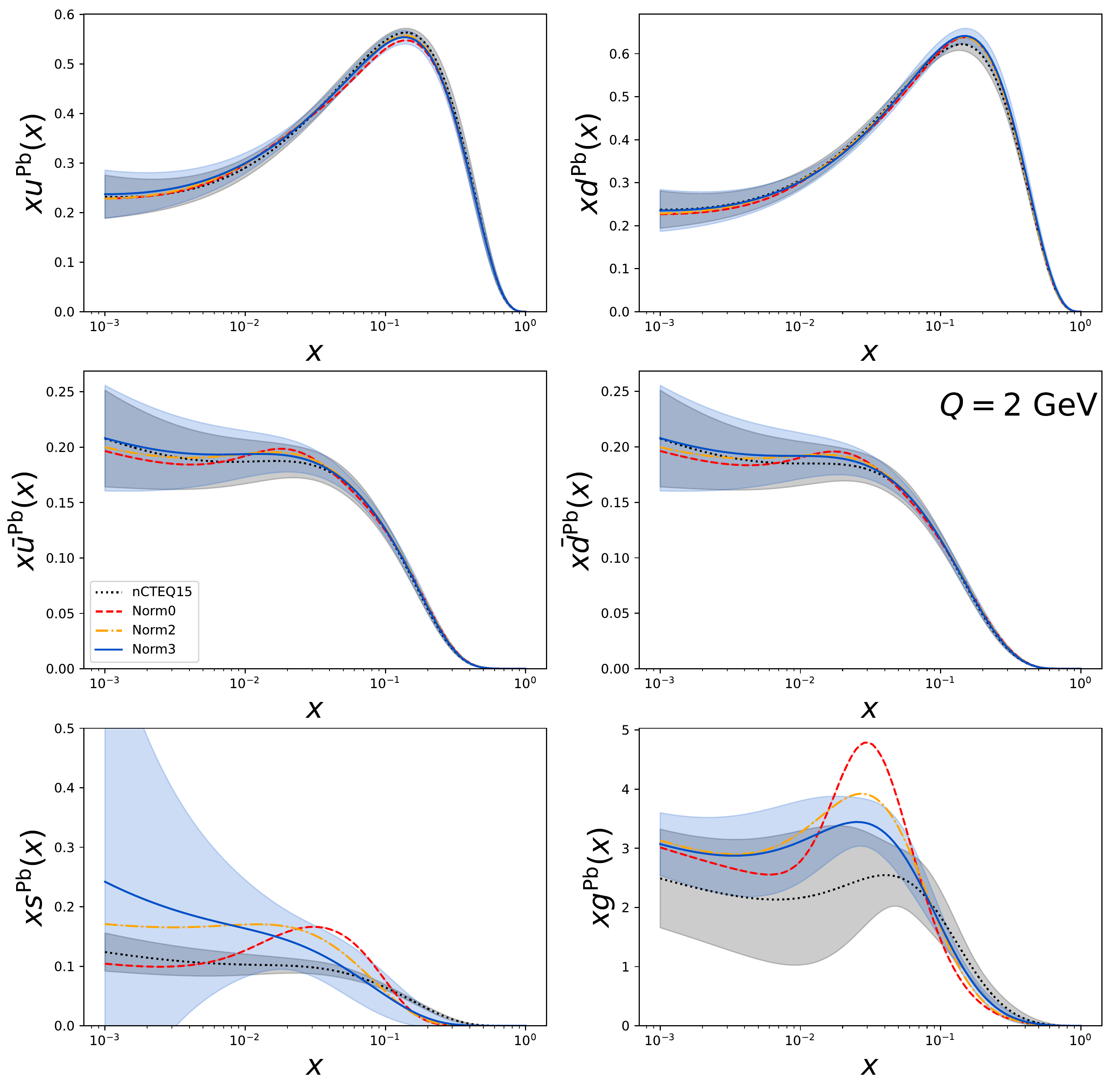}
\caption{The full  lead (Pb) PDFs for $Q=2$~GeV.
  The uncertainty band for nCTEQ15 is shown in gray,
  and for Norm3 in blue. 
The increase of the Norm0 set is evident for the strange
and gluon PDFs in the region of $x\sim 0.03$.}
\label{fig:fullNucPDFsQ2}
\end{center}
\end{figure*}
\begin{figure*}[tb]
\begin{center}
\includegraphics[width=\textwidth]{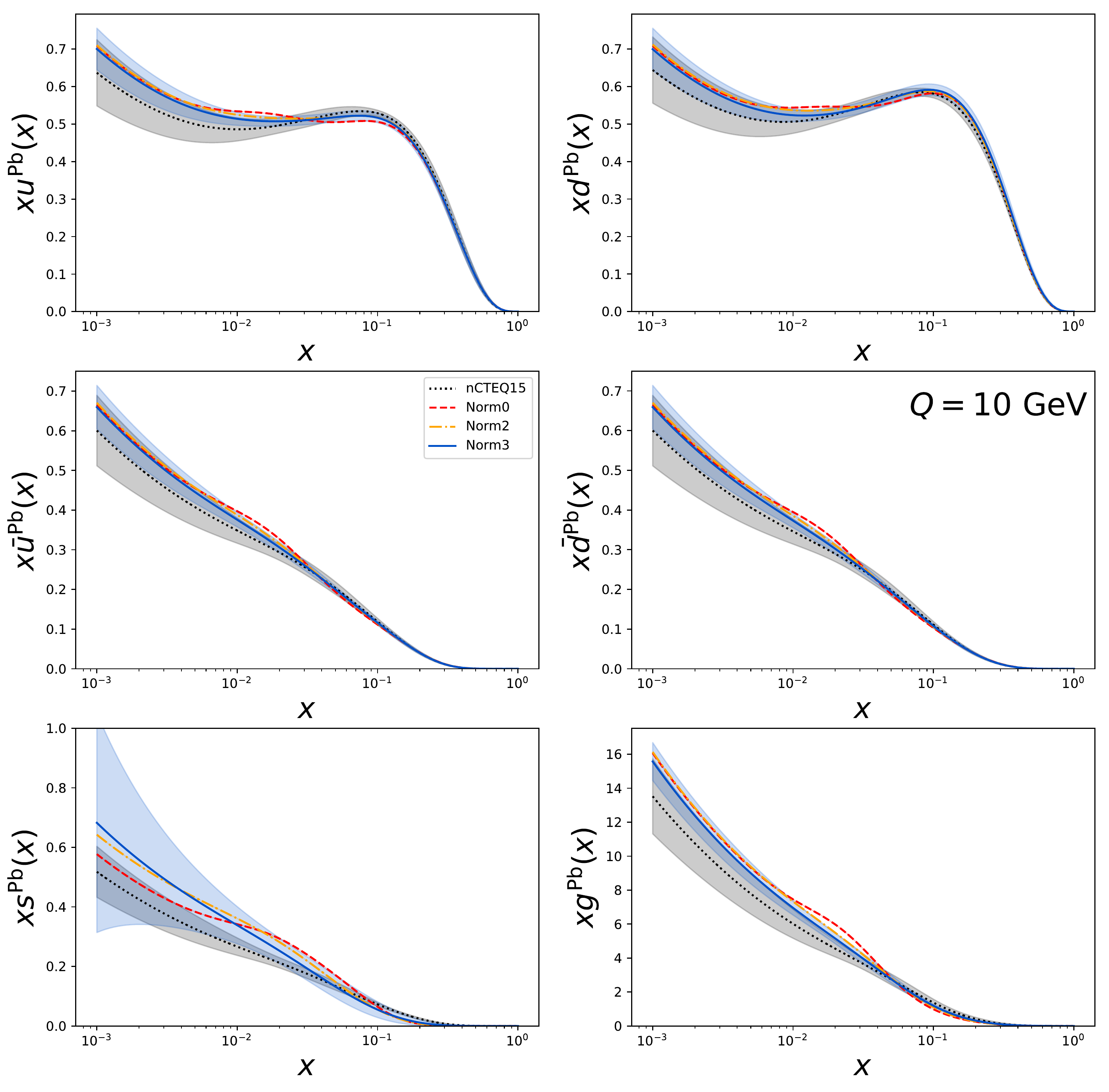}
\caption{The full  lead (Pb) PDFs for $Q=10$~GeV.
  The uncertainty band for nCTEQ15 is shown in gray,
  and for Norm3 in blue. 
  The increase in the Norm0 set  for the strange and gluon PDFs
  is reduced, compared to the lower $Q$ result,
  and shifted  to smaller  $x\sim 0.02$ values.}
\label{fig:fullNucPDFsQ10}
\end{center}
\end{figure*}
\begin{figure*}[tb]
\begin{center}
\includegraphics[width=\textwidth]{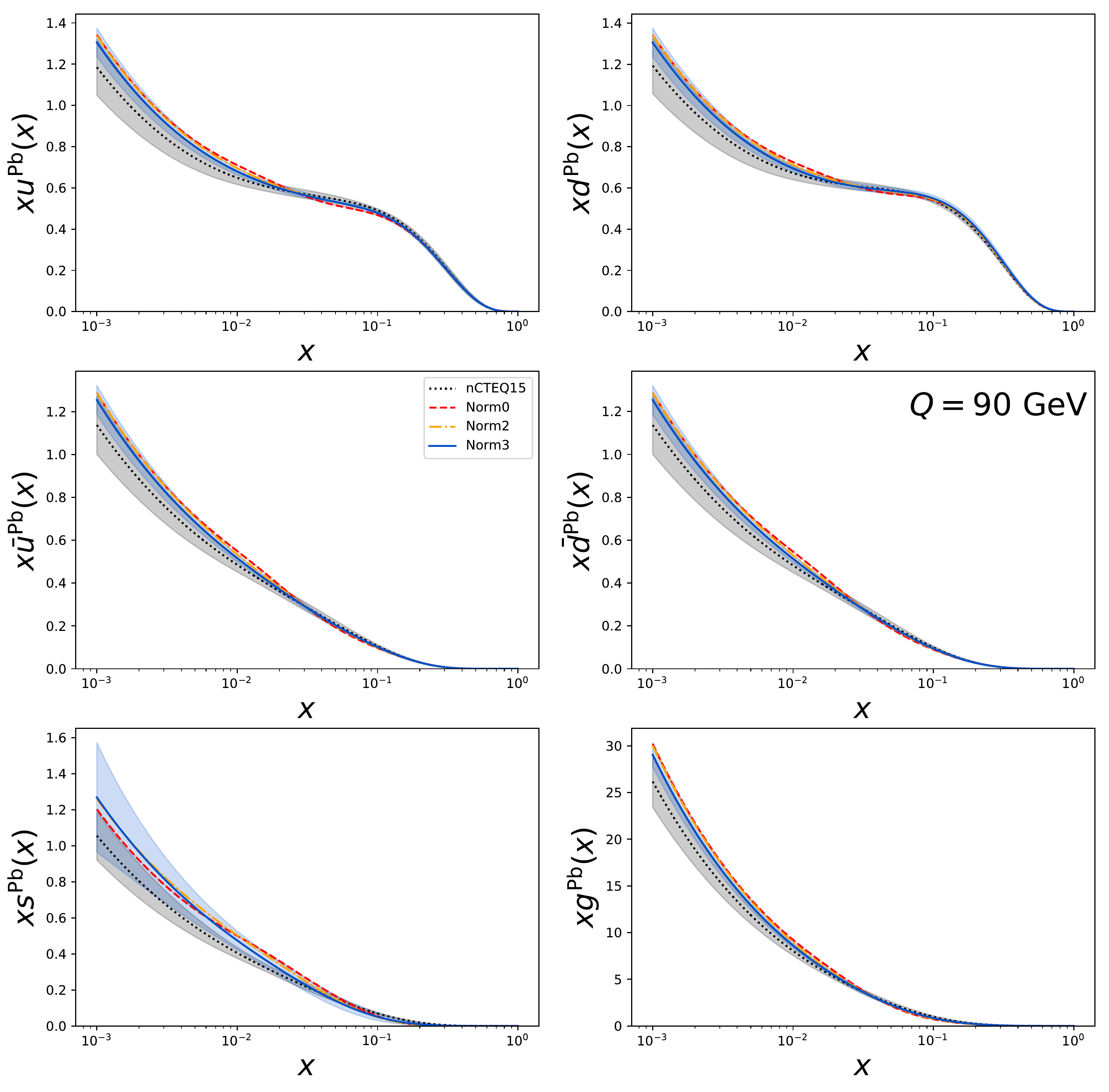}
\caption{The full lead (Pb) PDFs for $Q=90$~GeV.
  The uncertainty band for nCTEQ15 is shown in gray,
  and for Norm3 in blue. 
}
\label{fig:fullNucPDFsQ90}
\end{center}
\end{figure*}
\begin{figure*}[tb]
\begin{center}
\includegraphics[width=0.9\textwidth]{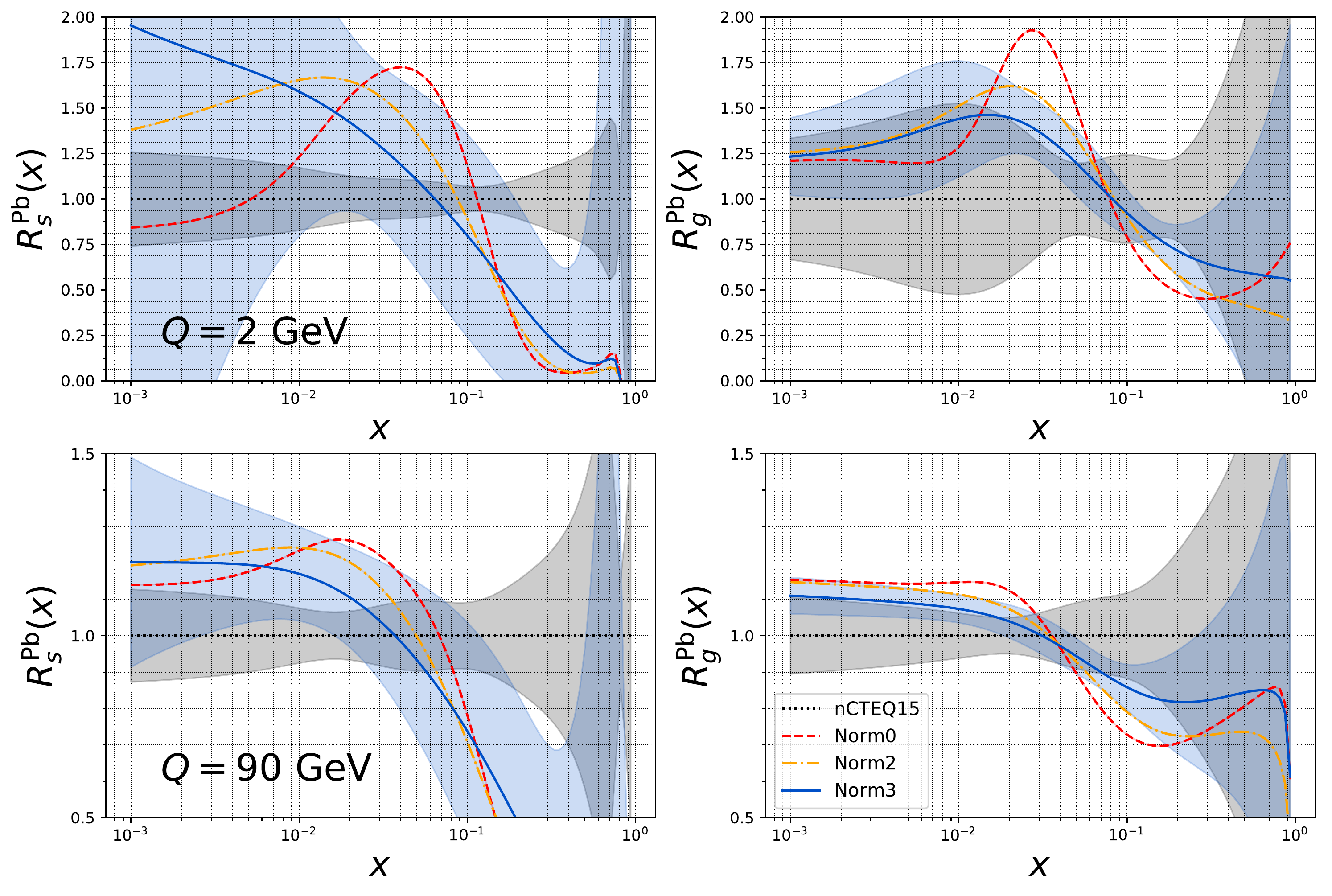}
\caption{Ratio of strange and gluon nPDFs compared to the corresponding
  nCTEQ15 nPDFs for $Q=2$ GeV (upper row) and $Q=90$ GeV (lower row).
  The uncertainty band for nCTEQ15 is shown in gray,
  and for Norm3 in blue. 
}
\label{fig:gsPDFratio}
\end{center}
\end{figure*}

Finally, we make a detailed examination of the
underlying flavor PDFs from these various fits. 
In Figs.~\ref{fig:fullNucPDFsQ2}--\ref{fig:fullNucPDFsQ90}
we display the nPDFs for a full lead nucleus at
three separate scales. 
The lowest scale ($Q=2$~GeV) is close to our initial evolution scale
of $Q_0=1.3$~GeV, the largest scale ($Q=90$~GeV) is in the range
relevant for $W^\pm /Z$ production, and the intermediate scale ($Q=10$~GeV)
helps illustrate the effects of the DGLAP evolution.

We choose to display the full lead nPDF as this is the physical quantity which enters the
calculation.\footnote{%
  Extracting a ``proton in a lead nucleus'' may introduce unphysical ambiguities
  when separating the up and down distributions.
  In particular, for an isoscalar target, we cannot separately
  distinguish the $u$ and $d$ distributions.}
This is computed using: 
\begin{equation}
  f_i^{(A,Z)} (x,Q) =
  \frac{Z}{A} f_i^{p/A} (x,Q)+
  \frac{A-Z}{A} f_i^{n/A} (x,Q) \quad ,
  \label{eq:fullnuc}
\end{equation} 
and we assume isospin symmetry to derive the neutron PDF.
\subsubsection{Strange and Gluon nPDFs:}

Examining the curves for up and down distributions, we see there is minimal
variation between different fits as these flavors are strongly constrained by other data.
Interestingly, we also see that the small~$x$ uncertainty is reduced at higher
scales (see Figs.~\ref{fig:fullNucPDFsQ10} and~\ref{fig:fullNucPDFsQ90}).
We observe a slight modification in the $\bar{u}$ and $\bar{d}$ distributions
as these are closely linked to the gluon and strange distributions which we will
discuss in the following. 

Turning to the gluon and strange PDFs, we see significant differences.
In particular, the fits seems to prefer a larger value for both the gluon and strange
PDFs  at intermediate $x$ values, which is the region relevant for the LHC heavy ion $W^\pm /Z$
production.
We discuss these fits in turn. 

\note{Norm0}
Examining  the Norm0 fit for $Q=2$~GeV (Fig.~\ref{fig:fullNucPDFsQ2}), 
we see a distinct excess in the strange and gluon PDFs in the region $x\sim 0.03$;
this is also evident in Fig.~\ref{fig:gsPDFratio} where we have plotted the
ratio relative to the nCTEQ15 values. 
At $Q=2$~GeV, the  peak of the gluon and strange distributions are
located at approximately $x\sim 0.03$;
via the DGLAP evolution these peaks shift down\footnote{%
  For comparison, in the ATLAS proton analysis,
  the central $x$ value at $\sqrt{s}=8$~TeV  corresponds to
  $M_{W/Z}/\sqrt{s}\sim 0.023$ at $Q_0=\sqrt{2}$~GeV,
  and evolves to  $x\sim 0.011$ at $Q\sim M_Z/\sqrt{s}$.
  }
 to the region  $x\sim 0.017$
for $Q=90$~GeV, consistent with the expectation for the central $x$ value of  $\sim M_{W,Z}/\sqrt{s}$.

Recall that the Norm0 fit does not allow any normalization adjustment in the fit.
Since the data  consistently lie above the theoretical predictions,
it appears that the Norm0 fit is exploiting the uncertainty of gluon and strange PDFs
to try and pull up the theoretical predictions in line with the data by increasing
the PDFs in the relevant $x$ region.
Additionally, we observe a similar (but less pronounced) behavior  in the $\bar{u}$ and  $\bar{d}$
distributions.%

As momentum must be conserved, we see the Norm0 strange PDF dips below nCTEQ15 at both
high and low $x$ values, while the gluon is below nCTEQ15 at higher $x$ values. 
Part of the reason the deformation of the gluon and strange PDFs is so large at  $Q=2$~GeV
is to compensate for the DGLAP evolution which will tend to diffuse the excess in the
gluon and strange distributions at the $Q=90$~GeV scale, {\it cf.}, Fig.~\ref{fig:fullNucPDFsQ90}. 
\note{Norm2 and Norm3}
In contrast to the Norm0 result above,
the Norm2 and Norm3 fits allow us to  investigate the effect of including the normalization
parameters into the fit; this is crucial in reducing the $\chidof$ for the LHC heavy ion data.
The effect on the resulting nPDFs is evident 
as shown in Fig.~\ref{fig:fullNucPDFsQ2} where we see that
the excess in both the strange and gluon is
systematically reduced as we introduce normalization parameters. 

In Fig.~\ref{fig:fullNucPDFsQ2} we also observe the greatly increased error band on
the Norm3 strange PDF as compared to nCTEQ15; this is of course due to
the additional fitting parameters for the strange quark included in the Norm3 analysis.

To highlight the magnitude of these differences, in Fig.~\ref{fig:gsPDFratio}
we plot the ratios of the PDFs compared to nCTEQ15.
At  $Q=2$~GeV, we see that the Norm0 gluon is nearly a factor of 2 times the nCTEQ15 value,
with a peak at $x\sim 0.03$.
The Norm2 and Norm3 gluon PDFs are reduced to  $\sim60\%$  and  $\sim40\%$
above nCTEQ15, respectively.\footnote{%
  Note we are focusing here on the intermediate $x$ region ($x \gtrsim
  0.01$) not only because this is the central $x$ region for \wz\ production,
  but because the small $x$ region is poorly constrained.}
Similarly, at $x\sim 0.03$ and  $Q=2$~GeV,
the strange PDF for both Norm0 and Norm2 are  $\sim60\%$ above the nCTEQ15 value,
while the Norm3 result is reduced to  $\sim25\%$.

In Fig.~\ref{fig:gsPDFratio} we also display ratio
at $Q=90$~GeV which illustrates the effect of the DGLAP evolution.
We see that the gluon is now reduced to $\sim 15\%$ above the nCTEQ15 value,
the strange is reduced to  $\sim 25\%$ above the nCTEQ15 value,
and both peaks have shifted to lower $x$ values.

Because the DGLAP evolution has ``washed out'' the detailed peak structure at low $Q$ values,
it is necessary for the fit to amplify the distortion at low $Q$ so that
a remnant of the effect survives at high $Q$.
Nevertheless, the remaining excess at $Q=90$~GeV is sufficient to improve the
$\chi^2$ of the fits. 

Additionally, we note that the heavy-flavor reweighting analysis of
Ref.~\cite{Kusina:2017gkz} also observed an increase of the gluon nPDF
in the intermediate to small $x$ region relative to the nCTEQ15 results.
While the shift of the PDF in the reweighting was in the same direction
as in this analysis, its magnitude was much smaller. 
We now turn our attention to the error band of the gluon distribution in
Fig.~\ref{fig:gsPDFratio}.
At NLO, the gluon enters for the first time the $W^{\pm}$ and $Z$ boson production
through the $gq$~initiated contributions.
The addition of the $W^{\pm}/Z$ LHC data to the fit is thus not expected to add
significant constraining power for the gluon distribution. Contrary to this
naive expectation,
due to high center of mass energy and relatively small values of the probed $x$,
the gluon distribution can have a considerable
contribution to $W^{\pm}/Z$ production processes; this is reflected in the
reduced error bands of Norm3 as compared with nCTEQ15. 
Indeed, an independent
variation of the open gluon parameters around the minimum in the Norm3 fit
confirms that the $\chi^2$ contribution from the LHC data is similarly steep
or steeper than contribution from all the other data included in the fit.\footnote{%
  In fact, this phenomenon is reminiscent of the significant impact of the gluon on the
  Tevatron high-$E_T$ jet cross sections via the $qg$-channel as described in Ref.~\cite{ Lai:1996mg}.}
\section{Comparisons  \label{sec:other}}

\subsection{Comparison with other nPDFs}

\begin{figure*}[tb]
\begin{center}
\includegraphics[width=\textwidth]{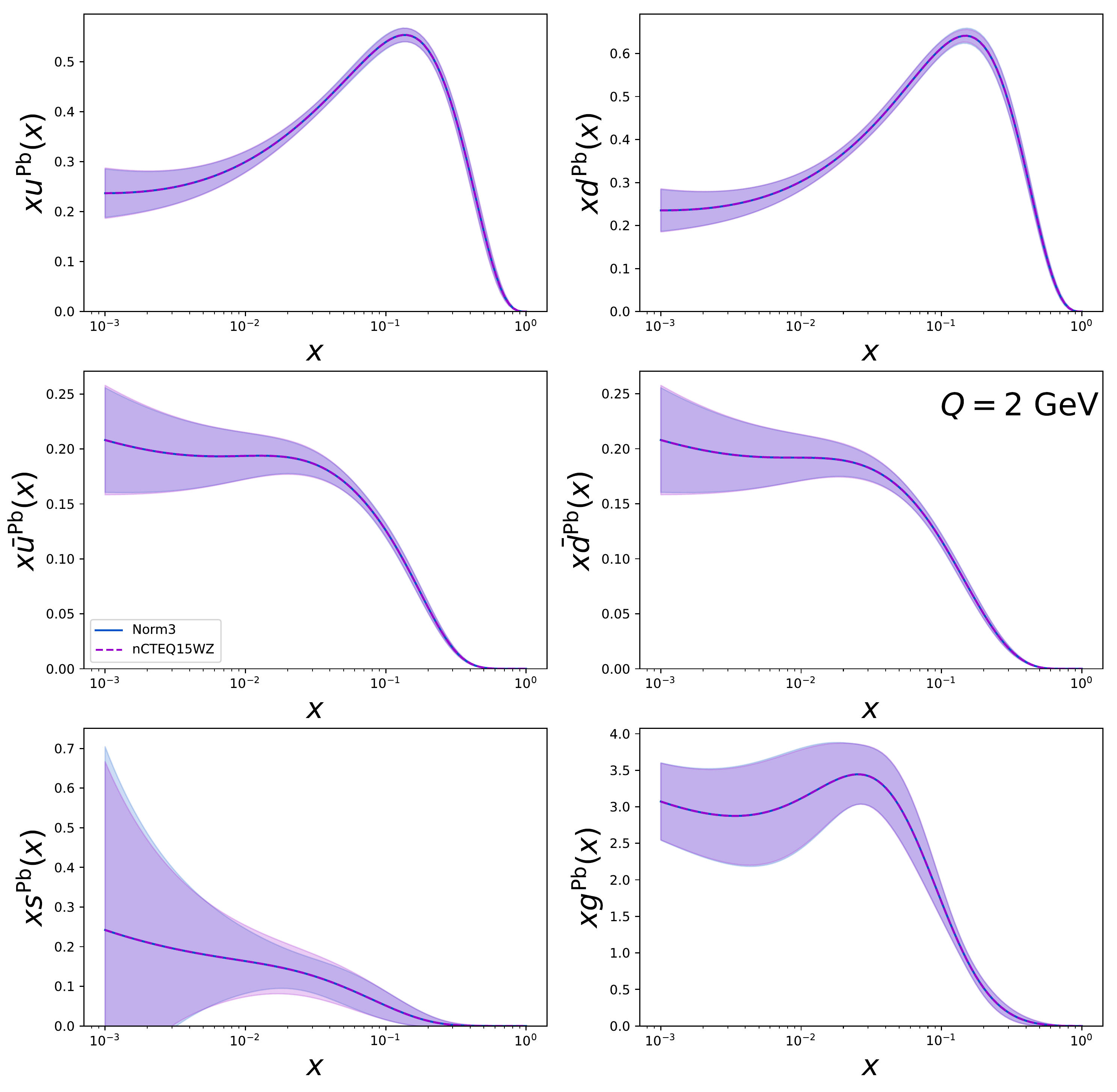}
\caption{
  Comparison of the full lead (Pb) PDFs at $Q=2$~GeV for
  nCTEQ15WZ and Norm3 fits.
  The uncertainty band for nCTEQ15WZ is shown in purple,
  Norm3 in blue.
  This shows that the central
  value of these fits are essentially identical, and the
  error bands are also virtually identical with the exception of small differences in the strange quark PDF.
} %
\label{fig:compare}
\end{center}
\end{figure*}
\begin{figure*}[tb]
\begin{center}
\includegraphics[width=\textwidth]{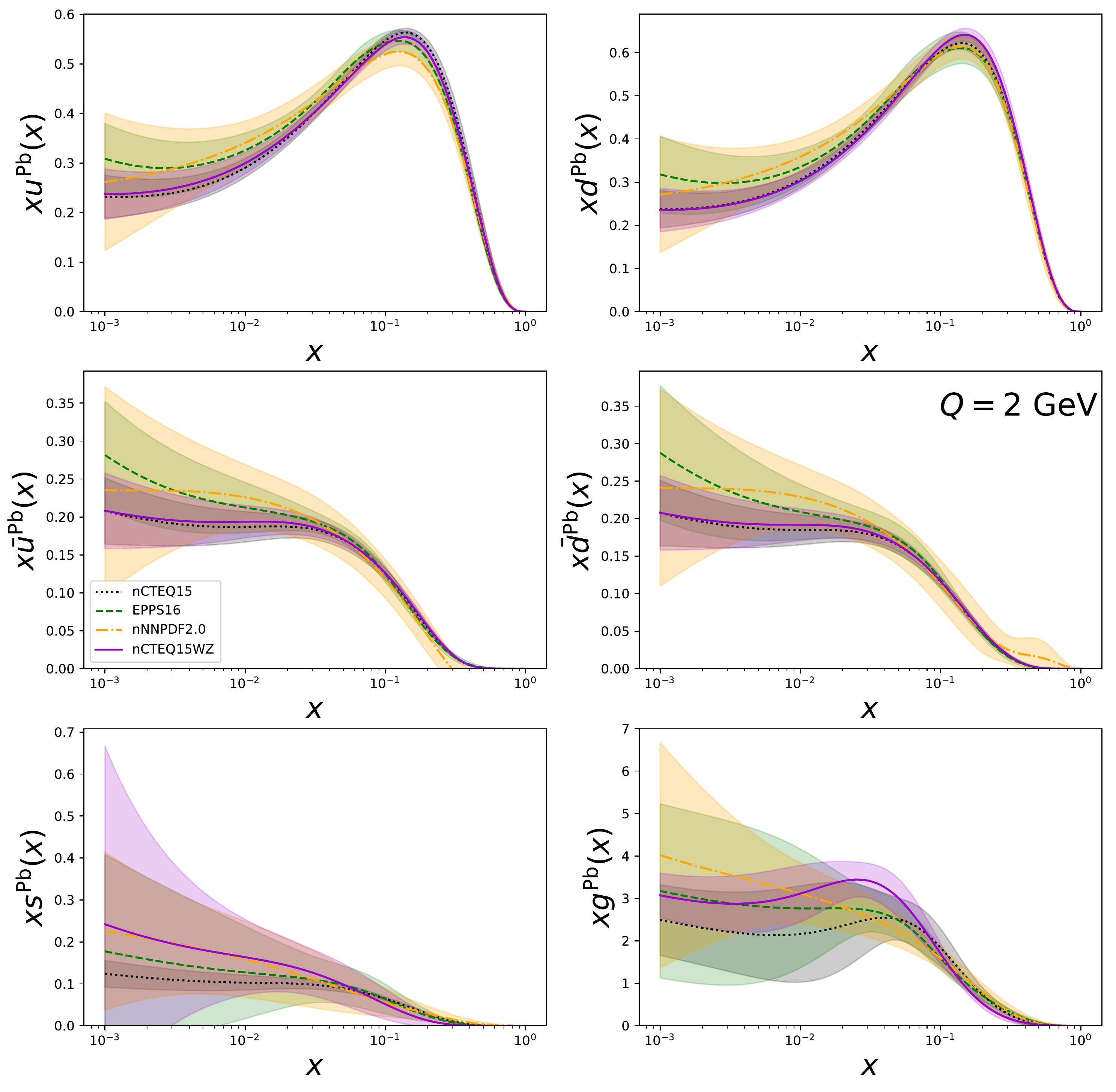}
\caption{
  Comparison of the full lead (Pb) PDFs at $Q=2$~GeV for
  nCTEQ15, EPPS16, nNNPDF2.0 and nCTEQ15WZ.
  The uncertainty band for nCTEQ15 is shown in gray,
  nCTEQ15WZ in violet, nNNPDF2.0 in yellow
  and EPPS16 in green.
} %
\label{fig:fullNucPDFsQ2EPPS}
\end{center}
\end{figure*}
\begin{figure*}[tb]
\begin{center}
\includegraphics[width=\textwidth]{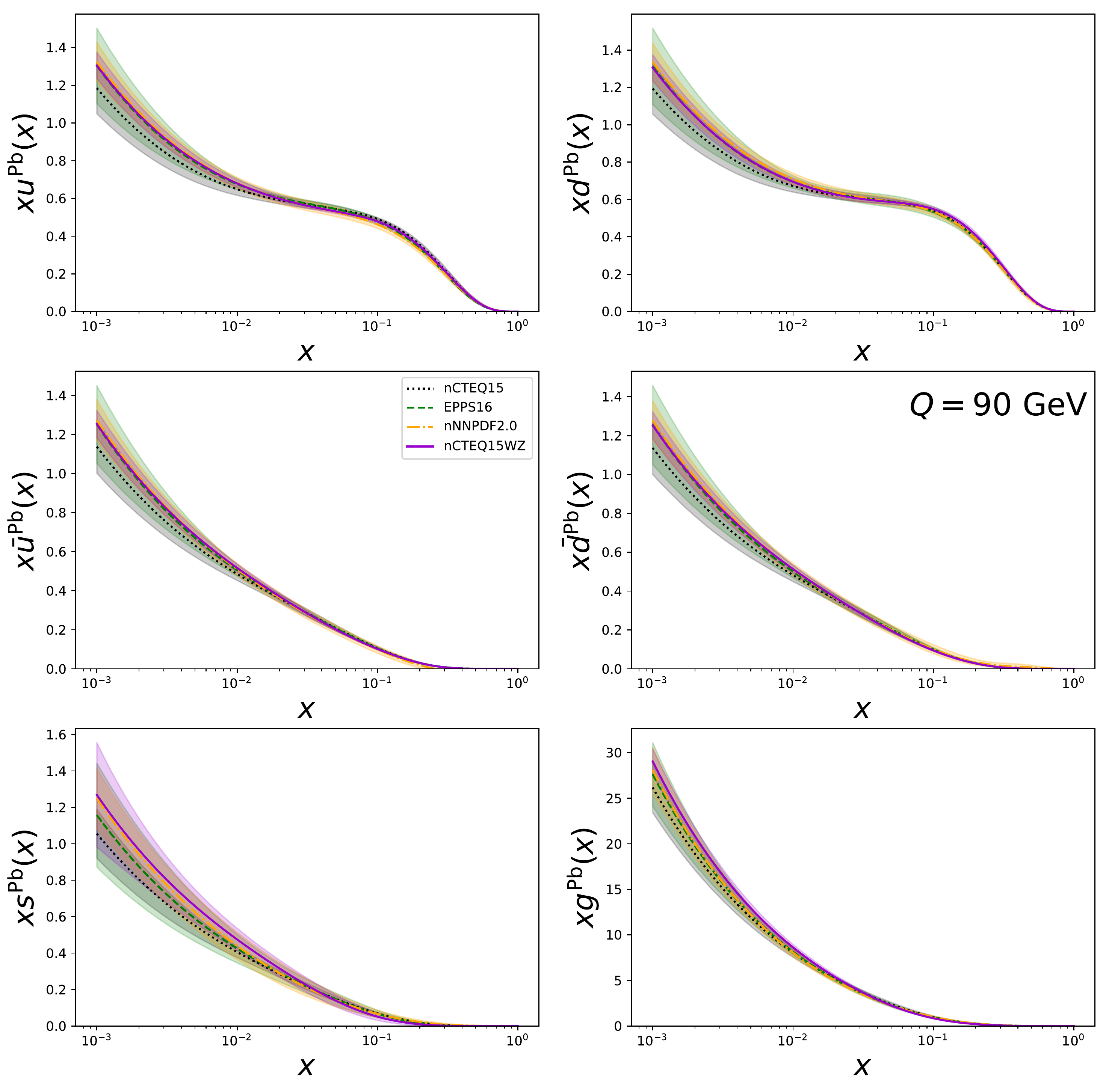}
\caption{
  Comparison of the full lead (Pb) PDFs at $Q=90$~GeV for
  nCTEQ15, EPPS16, nNNPDF2.0 and nCTEQ15WZ.
  The uncertainty band for nCTEQ15 is shown in gray,
  nCTEQ15WZ in violet, nNNPDF2.0 in yellow
  and EPPS16 in green.
}  %
\label{fig:fullNucPDFsQ90EPPS}
\end{center}
\end{figure*}

Having investigated the impact of the \wz\ heavy ion data including
normalization effects,
we now compare our PDFs with other results from the literature.

There are a number of nPDF sets
available~\cite{Hirai:2007sx,Eskola:2009uj,deFlorian:2011fp}
including some new determinations~\cite{Walt:2019slu,AbdulKhalek:2019mzd,AbdulKhalek:2020yuc}.
The TUJU19 analysis~\cite{Walt:2019slu} extends the  xFitter framework to include
nuclear PDFs; this open-source program provides a valuable tool for the PDF community. 
As an initial step, TUJU19 assumed $s=\bar{s}$ and $s=\bar{u}=\bar{d}$, and the
resulting nPDFs compare favorably with EPPS16 and nCTEQ15 within uncertainties.

A separate effort by the NNPDF collaboration~\cite{AbdulKhalek:2019mzd,AbdulKhalek:2020yuc}
uses neural network techniques to 
extract the gluon and quark  nPDFs; this method provides
a complementary approach to the traditional parameterized
function-based method.
Their recent analysis~\cite{AbdulKhalek:2020yuc} has produced the nNNPDF2.0 nPDF set
which includes  charged current DIS data from NuTeV (Fe) and Chorus (Pb),
and also LHC \wz\ data.
They also compute the  strangeness ratio, $R_s=(s+\bar{s})/(\bar{u}+\bar{d})$,
and find the nuclear value is reduced as compared to the proton.
The  neutrino DIS data and LHC $W+c$ associated production seem to prefer a lower $R_s$ value,
while the inclusive $W$ and $Z$ production favor a larger value.
These interesting observations raise some important issues, and 
additional investigation is warranted  to better understand the strange
distribution~\cite{Ethier:2020way}.

The EPPS16 data sets include DIS, DY, RHIC inclusive pion, and LHC \wz\ and dijet data;
in particular, this set incorporates a number of parameters to provide
flexibility in both the strange and gluon PDFs.
Therefore, it will be interesting to compare the variation of these
flavors between  our original nCTEQ15 nPDFs
and our nCTEQ15WZ fit. 

The nCTEQ15WZ fit is based on the Norm3 fit (with 3 normalization parameters),
and in addition includes the RHIC pion data in the fitting loop. 
The RHIC pion data is fit with the
Binnewies-Kniehl-Kramer (BKK) fragmentation functions~\cite{Binnewies:1994ju}
using a custom griding technique for 
fast evaluation~\cite{Kovarik:2015cma}.
The resulting nCTEQ15WZ nPDFs are nearly identical as the Norm3 nPDFs
which is evident when comparing the $\chidof$ values of Table~\ref{tab:Chi},
as well as the PDFs in Fig.~\ref{fig:compare}.

We now compare the results of our nCTEQ15WZ fit with the nCTEQ15, EPPS16,
and nNNPDF2.0 nPDFs in 
Figs.~\ref{fig:fullNucPDFsQ2EPPS} and~\ref{fig:fullNucPDFsQ90EPPS}.
To begin, we focus on the plots at $Q=2$~GeV as the variations are more evident here. 
For the up and down components $\{u, d, \bar{u}, \bar{d}\}$, nCTEQ15WZ
is quite similar to nCTEQ15, and these flavors generally lie below 
EPPS16 and nNNPDF2.0, but are within uncertainties.
For the strange and gluon, we see that nCTEQ15 and EPPS16 are generally similar for larger $x$ values,
and then diverge somewhat for small $x$.
The nCTEQ15WZ nPDFs lie below  nCTEQ15 and EPPS16 for large~$x$ values,
and then above at intermediate to small $x$ values; this allows $s(x)$ and $g(x)$
to increase the \wz\ cross section in the region of the data ($x\sim 0.02$)
while not perturbing the momentum sum rules.
nNNPDF2.0 is similar to  nCTEQ15 and EPPS16 for large~$x$ values,
but then increases for smaller $x$.
For the strange distribution, nNNPDF2.0 coincides
with nCTEQ15WZ at small $x$, while for the gluon,  nNNPDF2.0 exceeds nCTEQ15WZ at small $x$.
Similar effects to the above are generally evident at larger $Q$ values  (Fig.~\ref{fig:fullNucPDFsQ90EPPS}),
but their magnitude is diminished due to the DGLAP evolution effects.

\subsection{Comparison with proton results}

\begin{figure*}[tb]
\centering{}
\subfloat[]{
\includegraphics[width=0.45\textwidth]{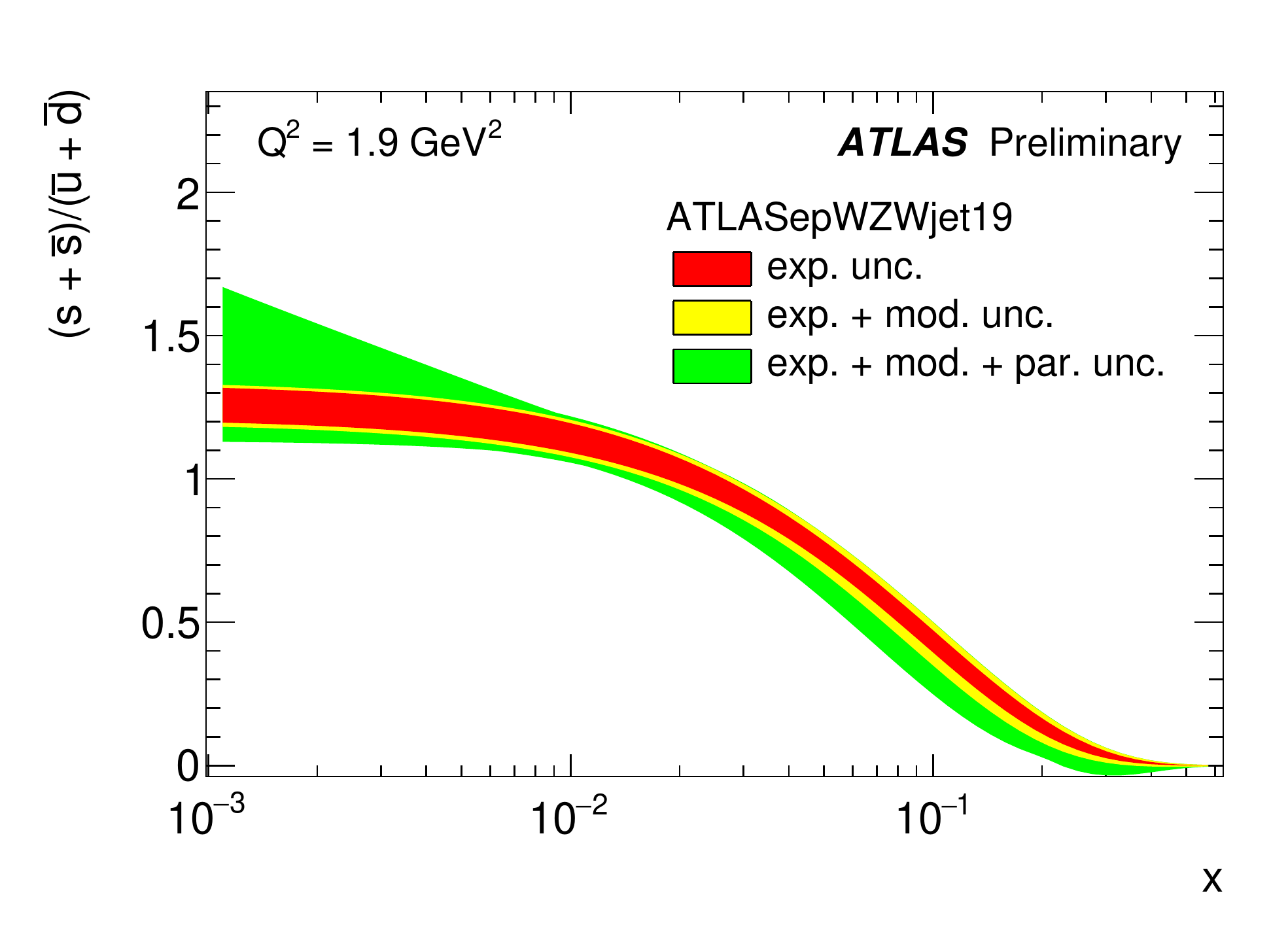}
\label{fig:rsATLAS}}
\hfil
\subfloat[]{
\includegraphics[width=0.45\textwidth]{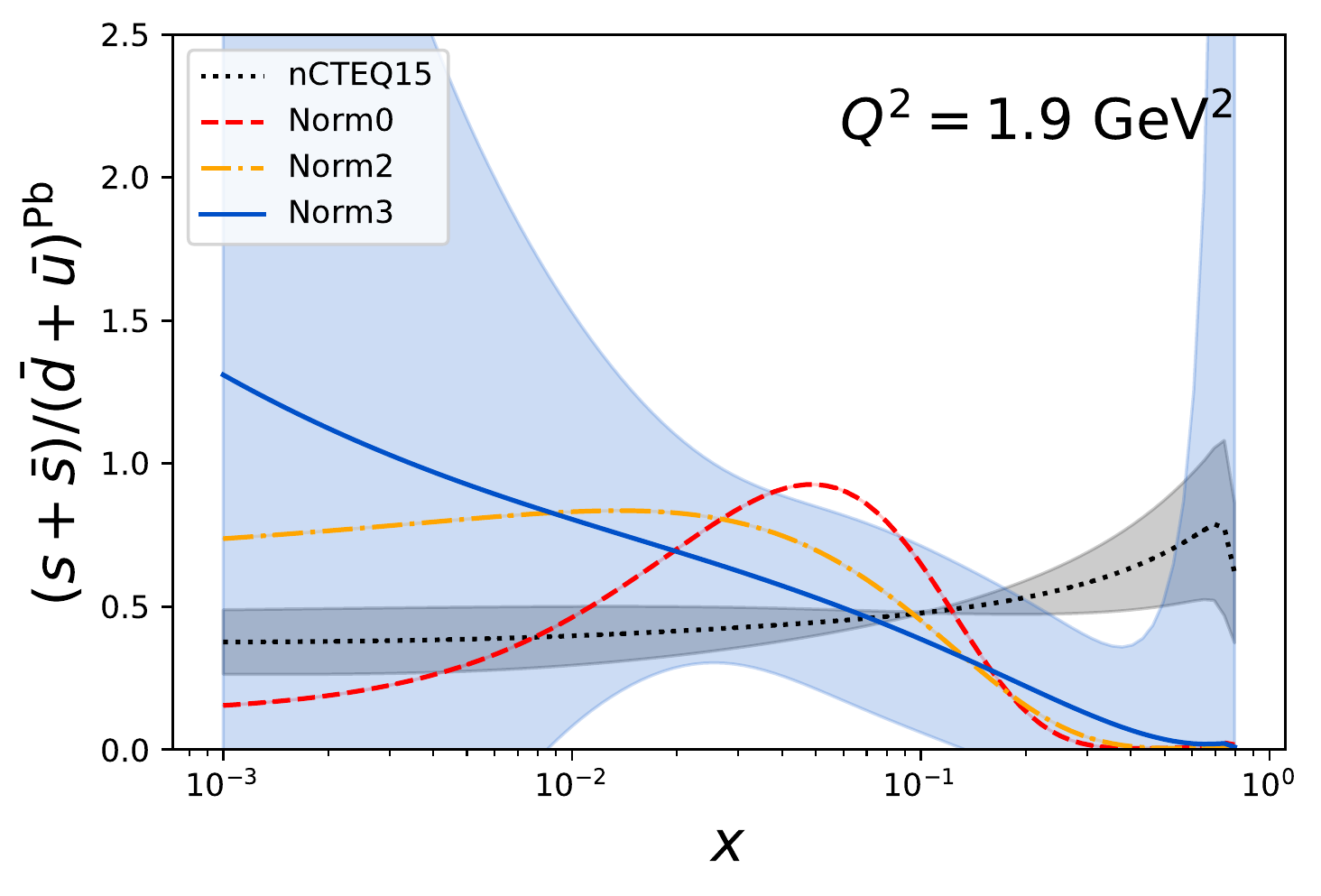}
\label{fig:rs}}
\\
\subfloat[]{
\includegraphics[width=0.45\textwidth]{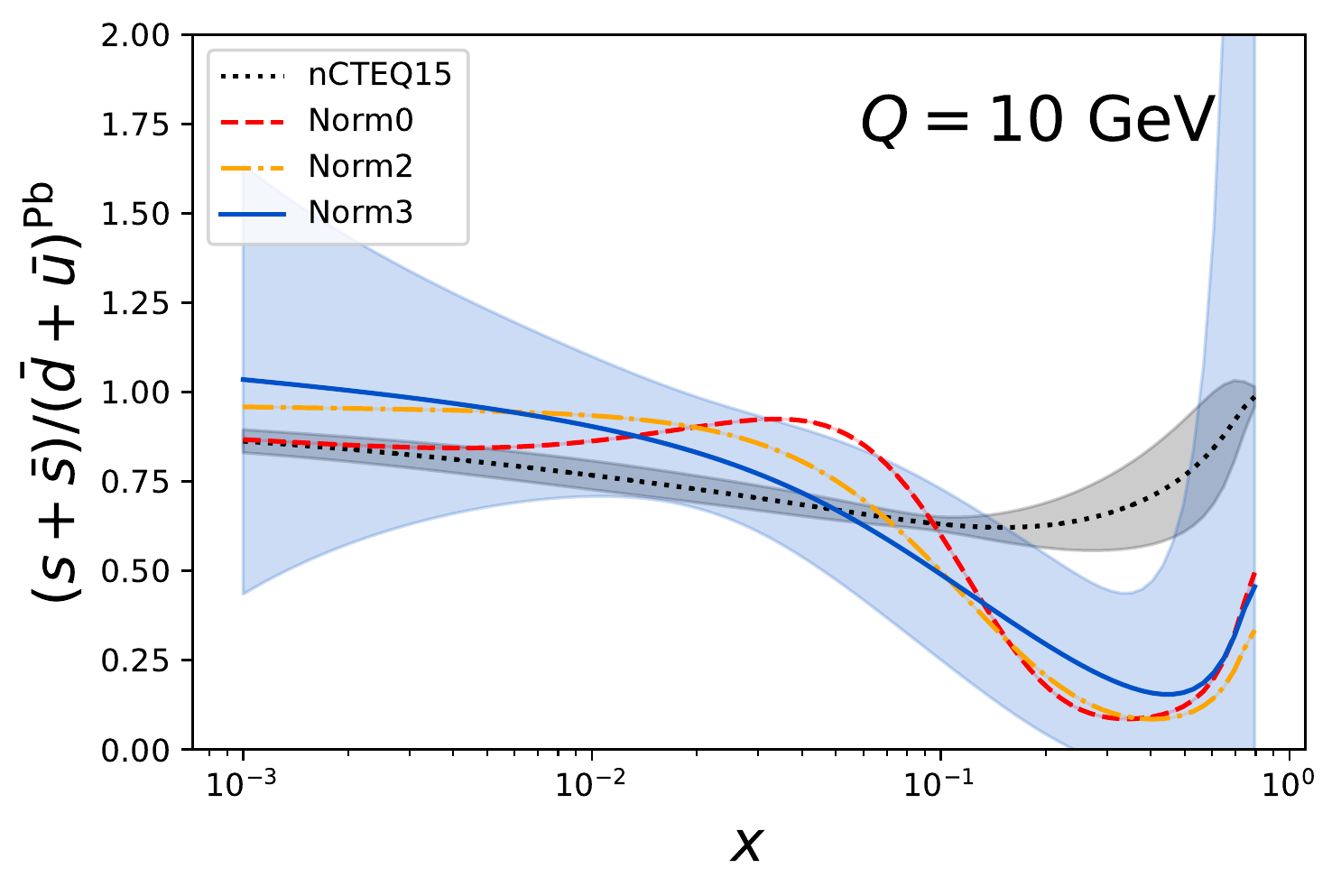}
\label{fig:rs10}}
\hfil
\subfloat[]{
\includegraphics[width=0.45\textwidth]{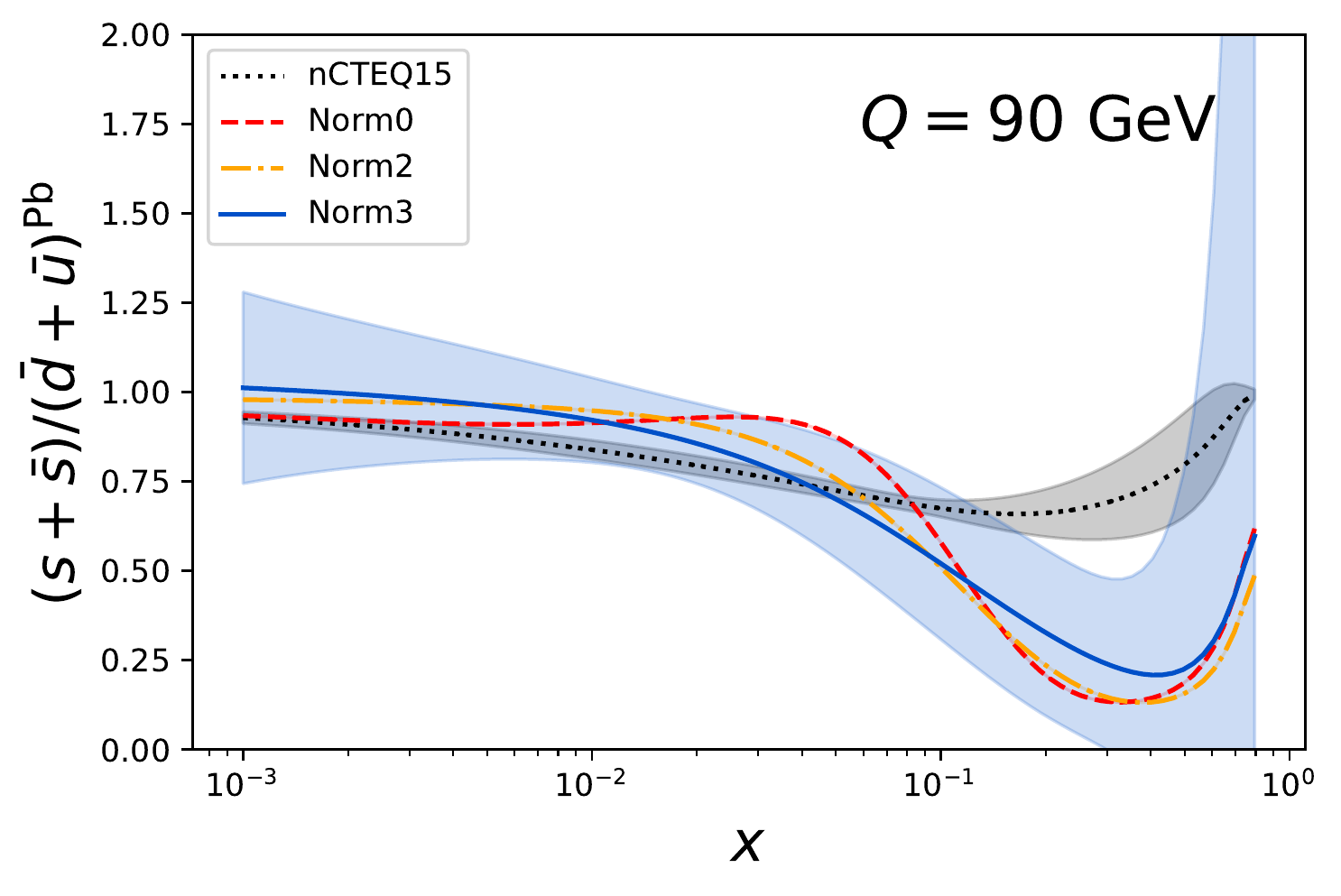}
\label{fig:rs90}}
\hfil
\caption{
  (a)~Recent (preliminary) result from ATLAS on the strange
  ratio for the proton~\cite{ATL-PHYS-PUB-2019-016}.
  \quad
  (b)~The nuclear strange ratio for lead (Pb) nPDFs as obtained in our fits.
  The uncertainty band for nCTEQ15 is shown in gray,
  and Norm3 in blue.
}
\label{fig:Rs}
\end{figure*}

The strange quark PDF has also been studied extensively for the proton case
by many groups including 
  ABM~\cite{Alekhin:2017olj},
  CT18~\cite{Hou:2019efy},
  JAM~\cite{Sato:2019yez},
  MMHT~\cite{Harland-Lang:2014zoa,Thorne:2019mpt},
  and 
  NNPDF~\cite{Ball:2009mk}.
  There is a close connection between the proton and nuclear PDFs;
  for example, nCTEQ15 uses the proton PDF as a boundary condition,
  and EPPS16 fits nuclear ratios relative to the proton.

One quantity of interest we can compare between the proton and the nuclear PDFs
is the  ratio of the  strange PDF relative to the light-sea quarks:
$R_s={(s+\bar{s})/(\bar{u}+\bar{d})}$.
In Fig.~\ref{fig:Rs},  we compute $R_s$ for selected $Q$ values,
and compare this to the proton result as extracted by
ATLAS~\cite{ATLAS:2019ext,Giuli:2019wtu}.

Comparing the proton and the lead results at $Q^2=1.9~{\rm GeV}^2$,
we see that the behavior of the  Norm3 curve  (\mbox{panel-b}) is
quite similar to the proton result (\mbox{panel-a}).
In contrast, the nCTEQ15 result is generally flat across all $x$ values
as the strange was set to be a fixed fraction of the $u/d$-sea PDFs,
$s=\bar{s}=\kappa(\bar{u}+\bar{d})/2$.
Additionally, we also display the other fits, Norm0 and Norm2,
to illustrate the range of possible variations.
The uncertainty bands for Norm3 are displayed;
these are large for small $x$, where the strange is poorly constrained,
and also at very large~$x$ where the quark sea denominator vanishes.
We also display larger $Q$ values which illustrates the convergent
effects of the DGLAP evolution.

In the previous section we raised the question as to
whether the enhanced strange distribution was reflecting the true underlying
physics, or was instead an artifact of the fit.
The similarities of $R_s$ between the proton and lead PDFs
may indicate that the
enhanced strange PDF is, in fact, a real effect. 
To definitively answer this question will require additional analysis, and this work is ongoing.

\goodbreak
\section{Conclusion \label{sec:conclusion}}

Our ability to fully characterize fundamental observables, 
like the Higgs boson couplings and the $W$ boson mass,
and to constrain both SM and BSM signatures is strongly limited by how
accurately we determine the underlying PDFs~\cite{Tanabashi:2018oca}. 
A precise determination of the strange PDF is an important step
in advancing these measurements.

The new nCTEQ++ framework allowed us to  include the LHC $W/Z$
data directly in the fit.  While these new fits significantly reduced the
overall $\chi^2$ for the $W/Z$ LHC data, we still observe tensions in individual
data sets which require further investigation.
Our analysis has identified factors which might further reduce
the apparent discrepancies  including: increasing the strange PDF,
modifying the nuclear correction, and adjusting the data normalization.

Compared to the nCTEQ15 PDFs, these new fits favor an increased
strange and gluon distribution in the $x$~region relevant for heavy ion  \wz{} production.
While we obtain a good fit in terms of the overall $\chi^2$ values,
we must ask: 
 i)~how the uncertainties and data normalization affect the resulting PDFs,
and
ii)~whether the results truly reflect the underlying physics,
or is the fit simply exploiting $s(x)$ because that is one
of the least constrained flavors?
The answer to this important question will require additional study;
this is currently under investigation.

\section*{Acknowledgments}

We are pleased to thank 
Aaron Angerami,
\'Emilien Chapon,
Cynthia Keppel,
Jorge Morfin,
Pavel Nadolsky,
Jeff Owens
and
Mark Sutton
for help and useful discussion.

A.K.\  is grateful for the support of the Kosciuszko Foundation.
A.K.\ also acknowledges partial support by Narodowe Centrum Nauki
grant UMO-2019/34/E/ST2/00186.
The work of T.J.\  was supported by the Deutsche Forschungsgemeinschaft
(DFG, German Research Foundation) under grant  396021762 - TRR 257.
The work of M.K.\  was funded by the Deutsche Forschungsgemeinschaft (DFG, German Research Foundation) -- Project-ID 273811115 -- SFB 1225.
T.J.H.\ and F.O.\   acknowledge  support through US DOE grant DE-SC0010129. 
T.J.H.~also acknowledges support from a JLab EIC Center Fellowship.

\appendix
\section{Fitting data normalizations}
\label{app:Norm}

When fitting the normalization of data sets, we use the $\chi^2$ prescription
given in Ref.~\cite{DAgostini:1993arp}.
For a data set $D$ with $N$ data
points and $S$ correlated systematic errors, the $\chi^2$ of the data set  reads:
\begin{eqnarray}
  \label{chi2}
    \chi^2_D &=& \sum_{i,j}^N \left(D_i - \frac{T_i}{N_{norm}}\right)
    (C^{-1})_{ij} \left(D_j-\frac{T_j}{N_{norm}}\right)
    \nonumber \\
    && \qquad + \left( \frac{1-N_{norm}}{\sigma_{norm}}\right)^2
\end{eqnarray}
where $\sigma_{norm}$ is the normalization uncertainty
and $T_i$ is the theoretical prediction for point $i$.
The last term of Eq.\eqref{chi2} is called the normalization
penalty and it enters when the fitted normalization, $N_{norm}$,
differs from unity.
The normalization uncertainty $\sigma_{norm}$ appearing in the denominator
prevents large excursions of  $N_{norm}$ away from unity.

The covariance matrix $C_{ij}$ is defined as:
\begin{equation}
  C_{ij} =
  \sigma_i^2 \, \delta_{ij}
  + \sum_\alpha^S \bar{\sigma}_{i\alpha}  \, \bar{\sigma}_{j\beta}
\end{equation}
where $\sigma_i$ is the total uncorrelated uncertainty (added in quadrature)
for data point $i$, and $\bar{\sigma}_{i\alpha}$ is the correlated systematic
uncertainty for data point $i$ from source $\alpha$.
Using the analytical formula for the inverse of the correlation matrix
as in Ref.~\cite{Stump:2001gu}, we obtain:
\begin{equation}
  \chi^2_D = \sum_i \left( \frac{D_i-T_i/N_{norm}}{\sigma_i}\right)^2
  - B^T A^{-1} B +\left( \frac{1-N_{norm}}{\sigma_{norm}}\right)^2,
\end{equation}
with
\begin{equation}
  A_{\alpha\gamma} = \delta_{\alpha \gamma}
  + \sum_i  \frac{\bar{\sigma}_{i\alpha} \, \bar{\sigma}_{i\gamma}}{\sigma_i^2},
\end{equation}
and
\begin{equation}
  B_\alpha = \sum_i\frac{\bar{\sigma}_{i\alpha} \, (D_i-T_i/N_{norm})}{\sigma_i^2}.
\end{equation}

\clearpage
\bibliographystyle{utphys}
\bibliography{refs,extra}
\end{document}